\documentclass[a4paper,12pt]{article}
\usepackage{amsmath}
\usepackage{amssymb}
\usepackage{amsfonts}
\usepackage{graphics}
\usepackage{epsfig}

\begin{document}
\renewcommand{\theequation}{\thesection.\arabic{equation}}
\begin{center}
{\Large\bf{Solving Noether's equations for gauge invariant local Lagrangians of N arbitrary higher even spin fields}}\\
\vspace{1cm}
{\large Werner R\"uhl}\\
Department of Physics, Technical University of Kaiserslautern \\
P.O. Box 3039,  67653 Kaiserslautern, Germany\\
\vspace{1cm}
\begin{abstract}
We consider systems of higher spin gauge fields that are described by a free field Lagrangian and one interaction of arbitrary order $N$ that is local and satisfies abelian gauge invariance. Such "solitary" interactions are derived from Noether potentials solving the Noether equations. They are constructed using a free conformal field theory carried by the same flat space as the higher spin fields. In this field theory we consider $N$-loop functions of conserved, conformally covariant currents, they are UV divergent. The residue of the first order pole in the dimensional regularisation approach to the $N$-loop function is a local differential operator and is free of anomalies, so that current conservation and conformal covariance is maintained. Applying this operator to the higher spin fields, the Noether potential results. We study the cases $N=2, N=3$ and $N=4$. We argue that our $N=3$ vertex for any number of derivatives $\Delta$ is identical with the known cubic interaction.
\end{abstract}
\vspace{2cm}
{\it February 2012}
\end{center}
\newpage

\section{Introduction}
In the investigations of higher spin gauge field theories that started with the seminal work of Fronsdal \cite{Frons}, the trilinear or cubic interactions of such fields played a central role. For small spins, i.e. two and three, Berends, Burgers and Van Dam  \cite{BBVD} constructed such interactions. In their work also the idea of finite or infinite series of $N$-linear interaction Lagrangians was put forward and this was connected with the concept of higher order gauge transformations, meaning that a $k$-th
order contribution to a gauge transformation depends on $k$ higher spin gauge fields locally, and on the gauge parameter $\epsilon$
linearly. Till today no such series has been constructed beyond $N=3$. We consider here only the gauge transformations for the case $k=0$, called "abelian".

After this classical work a long period of time was filled with investigations on the existence of general cubic interactions
\cite{Vas} and group theoretical analyses of criteria on the three spins which allow the interactions to exist \cite{Met}.
After these results became known, the first explicit examples of such interactions were produced, the examples, however, belong to
classes with special properties, e.g. conserved vector currents were constructed from two fields (such as the energy-momentum tensor) and these were coupled to the third field \cite{Boul, MMR1}. Finally explicit expressions for general cubic interaction Lagrangians were found \cite{MMR2,MMR3,MMR4}, and it was shown that the same expressions appear in the analysis of singular string models \cite{Sag}. Recent results are summarized in \cite{Mkrtchyan:2010pp}, \cite{Vasiliev:2011xf} and \cite{Taronna:2011kt}.

Both the cubic interactions ($N=3$) and the free Lagrangian ($N=2$) play an important role in the present investigation. Their explicit expressions are used to check (and understand) the results of the quantum field theory approach to any degree $N$ interaction Lagrangian with abelian gauge invariance obtained here. We prefer the notation $h^{(s)}(x;a)$ for the higher spin gauge field on the flat space with coordinates $x$ where, with an arbitrary tangent vector $a$ at $x$, the totally symmetric rank $s$ tensor field $h^{(s)}$ is homogeneous of degree $s$ in $a$. This notation allows to express all local operations on these tensor fields by differential operators in $x$ and $a$.

The results on cubic interactions as presented in \cite{MMR2,MMR3,MMR4} can be summarized as follows. The interaction Lagrangian consists of a sum of expressions $\mathcal{L}_{N=3}^{(i,j)}[h^{(s)}], 0  \leq i,j \leq 3, \, i+j \leq 3$ each of which is trilinear
in the fields
\begin{eqnarray}
h^{(s_{i})}(x;a) \qquad\qquad\qquad\qquad\nonumber\\
\bar h^{(s_{i})}(x;a) = \frac{1}{s_{i} (s_{i}-1)}\Box_{a}h^{(s_{i})}(x;a)\quad (\textnormal{trace field}) \nonumber\\
D h^{(s_{i})}(x;a) = \frac{1}{s_{i}}[(\partial_{a} \nabla) - \frac{1}{2} (a\nabla) \Box_{a}]h^{(s_{i})}(x;a)\quad (\textnormal{deDonder field})
\label{1.1}
\end{eqnarray}
and has the form
\begin{equation}
\mathcal{L}_{N=3}^{(i,j)} \sim \nabla^{k} (Dh^{(s_{\ell})})^{i} (\bar h^{(s_{m})})^{j} (h^{(s_{n})})^{3-j-i}
\label{1.2}
\end{equation}
where $\nabla^{k}$ denotes $k$ spacetime derivatives distributed in any way among the three types of fields, and the total number of derivatives  $k+i$ is (Metsaev's inequality \cite{Met})
\begin{equation}
\sum_{i=1}^{3} s_{i} -2\,\textnormal{min}\{s_{j}\} \leq k+i
\label{1.3}
\end{equation}
This total number is fixed for a Noether potential and denoted $\Delta$. Those expressions not containing deDonder or trace fields are denoted "leading" terms. From the derivation of the cubic interactions we learned several facts:
\begin{enumerate}
 \item For $i=0$ only the two cases $\mathcal{L}_{N=3}^{(0,j)}, j\in \{0,3\}$ are nonzero;
 \item Consequently in deDonder gauge, where $Dh^{(s)} = 0$, $h$ and $\bar h$ decouple as dynamical variables;
 \item "Proposition 1" is valid: All $\mathcal{L}_{N=3}^{(i,j)}$ can be reconstructed from the leading terms, they are unique.
\end{enumerate}
Proposition 1 follows from the construction technique used in \cite{MMR3}. Of course generalizations for arbitrary $N$ would be of great interest.

The strategy used in \cite{MMR2, MMR3, MMR4} has the following crucial aspects:
\begin{enumerate}
 \item In the coordinate representation the cubic interaction density is independent of the space dimension $D$. $D$ is a "hidden" parameter;
 \item The existence and value of Metsaev's minimal differential operator degree $\Delta_{min}$ \cite{Met} is taken as granted.
 For each $\Delta \geq \Delta_{min}$ a unique interaction Lagrangian (up to an overall normalization factor that may be chosen to    depend on $D$ of course) is obtained.
\end{enumerate}

The approach discussed in this work is based on a free scalar conformal quantum field theory carried by the same space as the higher spin fields. We construct conserved conformal currents each of an even spin $s$ from this scalar field, and evaluate the $N$-loop function of such currents. This approach has analogous aspects as the ones presented above from \cite{MMR2,MMR3,MMR4}. We have a nontrivial $D$ dependence of the loop graph function in momentum space, and its UV divergence is of course also $D$ dependent, so that dimensional regularization can be applied. In the resulting $\Gamma$ functions $D$-dependence
of the pole positions is obvious, and the sum of the residues which gives the singular part of the loop function we are interested
in, is $D$ dependent. But the final integration over the loop momentum $Q$ cancels the essential part of this dependence, namely up to its nonessential normalization factors (see (2.41), (2.42) and the comment following these equations). This allows us then to turn to part 2 of the strategy. We choose now $\Delta\geq \Delta_{min}$ freely as is done in \cite{MMR2,MMR3,MMR4}. In each case this has a consequence on the value of $D$. But part 1 of the strategy is interpreted as the permission to continue constant functions of $D$ "analytically" in $D$, allowing any integral value for $D$, even negative ones. We end up with the conclusion that for any $\Delta$ we find a unique cubic vertex function and this is supposed to be identical with that in \cite{MMR2, MMR3, MMR4}. However, only the case $\Delta = \Delta_{min}$ is studied in detail in Section 5. For all terms calculated in Section 5 this identity  between the results of both approaches has been verified. The construction gives a further result, namely that the vertex functions for any $N\geq 3$ are conformally invariant,
when conformal covariance of the higher spin functions is appropriately defined. For the case $N=2, \Delta =2$ the free Lagrangian is reproduced modulo trace terms (in the deDonder terms (1.1) the traces $\Box_{a}h^{(s)}$ must also be neglected). For $N=4$
we have also constructed the leading terms.

We define Noether potentials that are functionals of the higher spin gauge fields of the integral form
\begin{equation}
\Omega_{N}^{(K)}[h^{(s_{i})}] = \int dx \,\omega_{N}^{(K)}[h^{(s_{i})}](x)
\label{1.4}
\end{equation}
and are defined to be scalar polynomials of degree $N$ in derivatives of $K$ higher spin fields
and to obey
\begin{enumerate}
\item  locality,
\item  invariance with respect to abelian gauge transformations of all their functional variables $h^{(s_{i})}$ off shell.
\end{enumerate}
Here we consider a system of $K$ such fields
\begin{equation}
\{h^{(s_{i})}(z;a)\}_{i=1}^{K}
\label{1.5}
\end{equation}

The abelian gauge transformations are defined by
\begin{equation}
\delta_{0}h^{(s)}(x;a) = (a\nabla)\epsilon^{(s-1)}(x;a),\quad \Box_{a} \epsilon^{(s-1)}(x;a) = 0
\label{1.6}
\end{equation}
and abelian gauge invariance of the Noether potential $\Omega^{(K)}_{N}$ is expressed by
\begin{equation}
(\nabla\partial_{a})\frac{\delta\Omega^{(K)}_{N}[h]}{\delta h^{(s)}}(x;a) = 0
\label{1.7}
\end{equation}
for all $s$ and off-shell. These are the Noether equations. There are obviously $N$ of these, one for each field contained in $\Omega_{N}$. An interaction Lagrangian of degree $N$ denoted $\mathcal{L}_{N}[h]$ has the same locality properties but abelian gauge invariance is now fulfilled if the traceless part of
\begin{equation}
(\nabla\partial_{a})\frac{\delta \mathcal{L}_{N}[h]}{\delta h^{(s)}}\mid_{\textnormal{on-shell}}
\label{1.8}
\end{equation}
vanishes.

The Noetherian solving (1.7) has a polynomial form and is easily decomposed into the sum of two terms. The first vanishes on shell. We denote it the "kernel". The remainder is the candidate for the $N$-th order Lagrangian $\mathcal{L}_{N}[h^{(s_{i})}]$ if $N\geq3$. However, it may be useful to apply field redefinitions to this to obtain a standard form. Such field redefinitions are not investigated in this article.

In Section 2 we introduce the free scalar conformal quantum field theory which is used as a tool for our construction. The conserved currents of this one-field model are of particular interest. They have all even spin. Among these there are conformal currents and nonconformal derivatives of these. We argue that only the conformal currents play a role in our construction.
Moreover, in order to treat a higher spin field theory of Fronsdal's type, these fields are doubly traceless. The corresponding conformal currents should then be doubly traceless themselves, too, to eliminate unwanted extra interactions of the trace fields. Such currents are given in Section 2, eqs. (2.12), (2.14)-(2.16) and contain in the general case one free parameter multiplying a traceless current
inside the doubly traceless current. They can and must be employed in our algorithm as long as we consider the local covariant distribution defining the vertex.  However, if we contract this vertex distribution with external fields that are doubly traceless, the double and higher trace subtractions of the currents drop out, but the free parameter multiplying a simple trace current remains. With a trick we can show that the free parameter can also be transferred to the higher spin field where it multiplies a simple trace term of the field. So the QFT calculations can always be performed with the naive currents (2.2). The free parameter ("B-parameter") enters the vertex function again by means of the higher spin fields to which a simple trace term with this parameter is added at the end (see (2.18)). The result of the 2-vertex has identical leading terms with the free Fronsdal Lagrangian. On the other hand, what concerns the check of our algorithm for the $N=3$ vertex with the known result \cite{MMR2, MMR3, MMR4}, we study here only the case $\Delta_{min}$ and compare the leading terms and some divergence terms in both cases and find that they agree. Thus for the check on this level one can set the trace fields in the vertices for $N=2$ and $N=3$ equal zero by hand, and use the naive currents (2.2). We shall follow this strategy for $N=4$, too. Since Proposition 1 fixes unique expressions for trace and deDonder terms from the leading terms, (and suggesting that it holds also for $N=4$) there cannot remain a free parameter. We will argue later that the following Proposition 2 is valid: The $N$-point vertex functions and consequently the interaction Lagrangians for $N\geq3$ can be calculated by our algorithm with the naive currents (2.2).

Loop graphs for $N$ conserved conformal currents of arbitrary spin need a renormalization which is performed by dimensional regularization. The sum over the residues of all first order poles of the $\Gamma$-functions at positions $n\in \mathcal{N}_0$ (there is only a finite number of them) gives a local differential operator which produces the $N$'th order Noether potential. These residues being free of anomalies, current conservation and conformal covariance remain valid, and consequently the Noether equation is satisfied. The fact that the parameter $D$, the space-time dimension, drops out and $\Delta$ is fixed instead, implies that the counting label for the UV poles $n$ is no longer limited by an expression (2.35) containing $D$, but instead simply (see (2.37)) by
\begin{equation}
0\leq n \leq \frac{\Delta}{2}
\label{1.9}
\end{equation}
In Section 3 the parameter integrals that define the residues are further evaluated, leading to formulae to which we refer in the sequel.

Section 4 is devoted to the $N=2$ free Lagrangian, leading to a first check of our method. Section 5 is devoted to the vertex for $N=3$. As we emphasized, this is in principle a check of our quantum field theoretic ansatz. In Section 6 we study the case $N=4$. If $N>3$ typical complications arise due to the fact that $N$ momentum vectors submitted to momentum conservation possess bilinear forms that cannot be expressed linearly by momentum squares (Laplacians). This problem is dealt with in this article in a specific way, some elementary and technical relations are presented in an Appendix. The freedom of choosing this approach is equivalent to the possibility to act with partial integrations on the interaction Lagrangian density.

\section{The tools of the construction}
\subsection{The conserved currents}
\setcounter{equation}{0}
Given a free quantum field theory of a scalar massless field
\begin{equation}
\Box \Phi(x) = 0
\label{2.1}
\end{equation}
Spin $s$ currents are obtained for even $s$ from this scalar field by an operator product expansion
\begin{equation}
j^{(s)}(x;a) = (a,\nabla_1-\nabla_2)^{s} \Phi(x_1) \Phi(x_2)\mid{x_1=x_2=x}
\label{2.2}
\end{equation}
and their divergence is
\begin{eqnarray}
(\nabla_1+\nabla_2;\partial_{a})(a,\nabla_1-\nabla_2)^{s} \Phi(x_1)\Phi(x_2) = \nonumber\\ s((a,\nabla_1-\nabla_2)^{s-1}(\Box_1-\Box_2)\Phi(x_1)\Phi(x_2) = 0
\label{2.3}
\end{eqnarray}
since the free field $\Phi(x)$ satisfies the massless Klein-Gordon equation. For a related use made of massless free scalar fields and their conserved currents see \cite{Bekaert:2010ky}.

The conformal currents of spin $s$ are therefore all conserved. However, it is less well known that there are more conserved currents $j^{(s)}_{n}$ of spin $s$ which are derivatives of conformal operators but not all conformal themselves
\begin{equation}
j^{(s)}_{n}(x;a) = [(a^2\nabla^2 - (a,\nabla)^2]^{n} \times j^{(s-2n)}(x;a);\, n\in \{0,1,2,..s/2\}
\label{2.4}
\end{equation}
Conservation of these currents follows from
\begin{equation}
[(\nabla,\partial_{a}),(a^2\nabla^2 -(a,\nabla)^2)] = 0
\label{2.5}
\end{equation}
It follows then that the 2-point function of any pair of such different currents vanishes since
\begin{equation}
<0\mid j^{(s-2n_1)}(x_1;a_1) j^{(s-2n_2)}(x_2;a_2) \mid 0> = 0 \quad \textnormal{if}\quad n_1\not= n_2
\label{2.6}
\end{equation}
because of the inequality of the two conformal representations (spins and conformal dimensions are different).

We define the contraction of two symmetric tensors of equal spins by the star operation
\begin{equation}
j^{(s)}(x;a)\ast_{a}h^{(s)}(x;a) = (s!)^{-1}j^{(s)}(x;\partial_{a})h^{(s)}(x;a)
\label{2.7}
\end{equation}
and a generating function by the vacuum expectation value $<0\mid\Psi\mid 0>$ of
\begin{equation}
\Psi[j^{(s)}; h^{(s)}] = exp{\int d^{D}x j^{(s)}(x;a) \ast_{a} h^{(s)}(x;a) }
\label{2.8}
\end{equation}
where the vacuum belongs to the free conformal field theory of the scalar field operators $\Phi(x)$.
This generating function can be expanded in powers $N$ of $h$ starting with the trivial $1$ for $N=0$ and then a second order
($N=2$) term. Each power $N$ yields after regularization and extraction of the singular part the $N$'th order Noetherian.
Since the exponent of the generating function is second order in the free field $\Phi(x)$, the vacuum expectation value
of $\Psi$ is feasible. But it is unclear of how to regularize and extract the singular terms without using a series expansion
of the exponential function in advance. The vacuum expectation value of the current $j^{(s)}$ vanishes by definition (normal ordering).

If instead of the conformal current $j^{(s)}$ one would use the general conserved current
\begin{equation}
J^{(s)}(x;a) = \sum_{n=0}^{s/2} A_{n} j_{n}^{(s)}(x;a)
\label{2.9}
\end{equation}
one would destroy conformal covariance on the one hand, and on the other hand introduce factors of powers of
\begin{equation}
a^2p^2-(a,p)^2
\label{2.10}
\end{equation}
where $p$ is an external momentum, Laplacians and double divergences acting on the field $h^{(s)}$. In the interaction Lagrangians
there is no place for such terms (as known from the case $N=3$, \cite{MMR2}), they should be eliminated by field redefinitions if possible. In the case $N=2$ on the other hand the number of derivatives is restricted to $\Delta =2$. The orthogonality relation (2.6) requires that at least two such factors (2.10) appear if $A_{n} \not= 0$ for $n>0$, implying four and more derivatives. These are unwanted and therefore excluded. So it seems that the conformal covariant current is the only reasonable candidate for the generating function.

Since the current $j^{(s)}(x;a)$ in (2.8) is conformally covariant (spin $s$, dimension $D+s-2$) we can define conformal
covariance properties of $h^{(s)}(x;a)$ by the requirement of conformal invariance of the generating function, i.e. the representation of $h^{(s)}$ is dual to that of $j^{(s)}$. Then any unrenormalized $N$-loop function produced by the generating
function is also conformally invariant.

The conserved currents (2.2) possess nonvanishing traces denoted by
\begin{eqnarray}
J_0^{(s)}(x;a) = j^{(s)}(x;a) \qquad\qquad\qquad \nonumber\\
J_1^{(s-2)}(x;a) = \frac{\Box_{a}}{s(s-1)} j^{(s)}(x;a) \qquad\qquad \nonumber\\
J_2^{(s-4)}(x;a) = \frac{\Box^2_{a}}{s(s-1)(s-2)(s-3)}j^{(s)}(x;a)  \quad \textnormal{etc.}
\label{2.11}
\end{eqnarray}
Since the operators $\Box_{a}$ and $(\nabla,\partial_{a})$ commute, conservation and vanishing of multiple traces are compatible requirements. Therefore starting from the conserved currents $j^{(s)}(x;a)$ we can construct multiple traceless conserved currents by the expansion ("subtraction of traces", $B_0 = 1$)
\begin{equation}
\tilde J^{(s)}(x;a) = \sum_{n\geq0} B_{n}^{(s)}\, (a^2)^{n} J_{n}^{(s-2n)}(x;a)
\label{2.12}
\end{equation}
Requiring the single trace of $\tilde J^{(s)}(x;a)$ to vanish leads to \footnote{We use the notation $[x]_{n} = \prod_{k=0}^{n-1}(x-k)$ and $\mu =D/2,\,D$ the space dimension}
\begin{equation}
B_{n}^{(s)}(single) = (-1)^{n} \frac{[s]_{2n}}{2^{2n}n! [s+\mu-2]_{n}},\quad
\label{2.13}
\end{equation}
If the double trace of $\tilde J^{(s)}(x;a)$ is required to vanish, we use $B_1^{(s)} = B_1^{(s)}(double)$ as a free parameter and expand
\begin{equation}
B_{n}^{(s)}(double) = \alpha_{n}^{(s)} + B_1^{(s)} \beta_{n}^{(s)},\quad \alpha_0^{(s)} = \beta_1^{(s)} = 1,\quad \alpha_1^{(s)} = \beta_0^{(s)} = 0
\label{2.14}
\end{equation}
Solving the recursions in this case yields
\begin{eqnarray}
\alpha_{n}^{(s)}=  (-1)^{n-1}\frac{[s]_{2n}}{2^{2n} n(n-2)![s+\mu-3]_{n}} , \quad n\geq 2 \label{2.15} \\
\beta_{n}^{(s)} =  (-1)^{n-1}\frac{[s-2]_{2n-2}}{2^{2n-2}(n-1)! [s+\mu-4]_{n-1}}, \quad  n\geq 1 \label{2.16}
\end{eqnarray}
If in (2.7) we replace $j^{(s)}(x;a)$ by either of these choices $\tilde J^{(s)}(x;a)$, then only that part of the field $h^{(s)}(x;a)$ which is traceless (respectively is doubly traceless) is coupled to the current. By comparison of (2.16) with (2.13)
we find that
\begin{equation}
\beta_{n}^{(s)} = B_{n-1}^{(s-2)}(single)
\label{2.17}
\end{equation}
belonging to a simply traceless current. Though (1.8) expresses that the divergence of the traceless part of the current must vanish, this requirement is trivial if the total current is conserved. Thus $B_1^{(s)}$ remains a free parameter indeed.
Of course it is also trivial that any doubly traceless current remains doubly traceless if an arbitrary simply traceless current is added.

Contracting the current $\tilde J^{(s)}(x;a)$ with the doubly traceless higher spin field $h^{(s)}(x;a)$ by the $\ast_{a}$-operation (2.7) gives
\begin{eqnarray}
\{j^{(s)}(x;a) + \frac{B_1^{(s)}} {[s]_2}a^2 (\Box_{a}j)^{(s-2)}(x;a) \}\ast_{a}\, h^{(s)}(x;a) =\nonumber\\
j^{(s)}(x;a) \ast_{a} \{h^{(s)}(x;a) +\frac{B_1^{(s)}}{[s]_2}a^2(\Box_{a}h)^{(s-2)}(x;a)\}
\label{2.18}
\end{eqnarray}
Consequently in all our formulas for the vertex functions we can evaluate N-point loop functions as vacuum expectation values of the currents $j^{(s)}(x;a)$ (2.2) and multiply these loop functions not with $h^{(s)}$ but with $h^{(s)} +B_1^{(s)}a^2 Tr h^{(s-2)}$. In this article we calculate all terms including trace and deDonder terms only for $N=2$, the leading plus some divergence terms for $N=3$ and only the leading terms for $N=4$. In this context we refer again to the Proposition 1 formulated in Section 1. Its validity for $N\geq 4$ has not been proved yet, however, but it could be true in general.

\subsection{The N-point loop functions}
In the Hilbert space of the free field $\Phi(x)$ over the $D$ dimensional flat space we consider the connected $N$-point function
\begin{equation}
<0\mid \prod_{i=1}^{N} j^{(s_{i})}(x_{i};a_{i})\mid0>_{conn}
\label{2.19}
\end{equation}
which consists of
\begin{equation}
\frac{1}{2}(N-1)!
\label{2.20}
\end{equation}
contractions (by Wick's theorem) e.g. of the type
\begin{equation}
\prod_{i=1}^{N}(a_{i}\nabla_{i})^{s_{i}-r_{i}} (a_{i+1}\nabla_{i+1})^{r_{i+1}} ((x_{i} - x_{i+1})^2)^{-(\mu-1)},\qquad \mu = \frac{D}{2}
\label{2.21}
\end{equation}
We restrict our discussion to this example. All other examples would be mathematically equivalent.

Going to the momentum representation
\begin{eqnarray}
\int \prod_{j=1}^{N}dx_{j}\exp\{ix_{j}p_{j}\},\qquad \int \prod_{j=1}^{N}dx_{j} \exp\{i(x_{j}-x_{j+1})q_{j,j+1}\} \nonumber\\
\nabla_{i} \rightarrow p_{i} = q_{i,i+1} - q_{i-1,i}\qquad\qquad
\label{2.22}
\end{eqnarray}
and neglecting constant and only $D$-dependent factors, and postponing a factor $(-1)^{\sum r_{i}}$ we have the integral
\begin{eqnarray}
\int \prod_{i=1}^{N}dq_{i,i+1} (q_{i,i+1}^2)^{-1} \prod_{j=1}^{N}(a_{j}q_{j,j+1})^{s_{j}-r_{j}}\prod_{k=1}^{N}(a_{k}q_{k-1,k})^{r_{k}}\nonumber\\
\prod_{\ell=1}^{N}\delta(p_{\ell}-q_{\ell,\ell+1}+q_{\ell-1,\ell})\qquad\qquad
\label{2.23}
\end{eqnarray}
Here labels $1\leq i \leq N$ are used cyclically, so that $\ell+1$ for $\ell =N$ means $1$ and $k-1$ for $k=1$ means $N$. Moreover from the Fourier integrals imaginary units $i$ result which multiply to $(-1)^{R}\, i^{S}, R = \sum_{i}r_{i}, S = \sum_{i}s_{i}$ that cancel the $r$-dependent signs from all expansion coefficients $A_{r_{i}}^{(s_{i})} = (-1)^{r_{i}}{s_{i} \choose r_{i}}$ in (2.2).

These delta-functions can be taken into account by solving for the $q_{i,i+1}$ in terms of the $p_{i}$ and a new momentum $Q$
\begin{equation}
q_{i,i+1} = Q +(p_1+p_2+...+p_{i}) = Q +w_{i}, \quad  q_{N,1} = Q, \quad \sum_{i=1}^{N} p_{i} =w_{N} = 0
\label{2.24}
\end{equation}
The resulting integral is
\begin{equation}
\int dQ [Q^2 (Q+w_1)^2 (Q+w_2)^2 ...(Q+w_{N-1})^2]^{-1} \prod_{i=1}^{N}(a_{i}(Q+w_{i}))^{s_{i}-r_{i}}(a_{i}(Q+w_{i-1}))^{r_{i}}
\label{2.25}
\end{equation}
The square bracket in the integral can be represented by
\begin{equation}
\int_{0}^{\infty} \prod_{i}^{N} d\sigma_{i} \exp\{-[\sigma_1 Q^2 +\sigma_2 (Q+w_1)^2+\sigma_3 (Q+w_2)^2 +...+ \sigma_{N}(Q+w_{N-1})^2]\}\times...
\label{2.26}
\end{equation}
We introduce
\begin{equation}
\Sigma = \sum_{i=1}^{N}\sigma_{i}, \quad \frac{\sigma_{i}}{\Sigma} = \tau_{i}
\label{2.27}
\end{equation}
so that the exponent can be rewritten as
\begin{equation}
\{-\Sigma[(Q+\tau_2w_1 +\tau_3w_2+ ... +\tau_{N}w_{N-1})^2+ Z(\tau, w)]\}
\label{2.28}
\end{equation}
Now denote
\begin{equation}
Q' = Q +\sum_{i=1}^{N-1}\tau_{i+1}w_{i} = Q + \chi(\tau,w)
\label{2.29}
\end{equation}
and introduce the integration variable $Q'$ instead of $Q$.

This must be done also in the gradient terms after multiplication with $\mid A_{r_{i}}^{(s_{i})}\mid$ (now the sign factor from (2.23) is taken into account) and summing
\begin{equation}
\sum_{\{r_{i}\}}\prod_{i=1}^{N} {s_{i} \choose r_{i}} (a_{i}(Q+w_{i})^{s_{i}-r_{i}}(a_{i}(Q+w_{i-1})^{r_{i}} =
\prod_{i=1}^{N} (a_{i}(2Q+w_{i}+w_{i-1}))^{s_{i}}
\label{2.30}
\end{equation}
and then inserting $Q'$
\begin{eqnarray}
2Q + w_{i}+ w_{i} = 2Q' + 2R_{i}, \quad R_{i}(\tau,w) =\frac{1}{2}(w_{i}+w_{i-1}) - \chi(\tau, w)\qquad\qquad \label{2.31}\\
(a_{i},Q' + R_{i})^{s_{i}} = \sum_{m_{i}}{s_{i}\choose m_{i}}(a_{i}Q')^{m_{i}}(a_{i}R_{i})^{s_{i}-m_{i}}\qquad \label{2.32}
\end{eqnarray}
In the sequel we skip the factor $2^{S}, S = \sum_{i} s_{i}$.

Now we rescale $Q'$ by $Q' = \Sigma^{-\frac{1}{2}}Q''$ and obtain after $Q''$ integration
\begin{eqnarray}
\int_{0}^{\infty} d\Sigma \,\Sigma^{N-1 -\frac{1}{2}(M+D)}\exp\{-\Sigma Z \}  \nonumber\\
= \Gamma(N-\frac{1}{2}(M+D)) Z^{-N+\frac{1}{2}(M+D)} \label{2.33}\\
M = \sum_{i=1}^{N} m_{i}, \qquad\qquad \label{2.34}
\end{eqnarray}
In order to obtain a local vertex,
\begin{equation}
n = \frac{1}{2}(M+D) -N
\label{2.35}
\end{equation}
must be natural nonnegative, because this and only this guarantees that this result is polynomial in $\{w_{i}\}$.
Therefore we must have an ultraviolet divergence from the gamma function. In its argument we replace $D\rightarrow D -2\epsilon$
leading to a pole behaviour
\begin{equation}
\Gamma(N-\frac{1}{2}(M+D)) = \frac{(-1)^{n}}{n!} \epsilon^{-1} + O(1)
\label{2.36}
\end{equation}
and we have to collect all terms with first order pole in $\epsilon$ and sum over their residues. Since $Z$ is of second order in
the momenta (derivatives), $2n$ is the number of derivatives contributed by $Z^{n}$. The power of momenta contributed by the second factors on the r.h.s. of (2.32) is $S-M$. Thus in order to obtain a fixed number $\Delta = S-M+2n$ of derivatives of the whole
expression we must allow an $n$-dependence of $M$: replace $M$ by $M_{n} = M_{0} +2n$, and correspondingly we use $\Delta_{n} = S-M_{n} = \Delta -2n$. From the last equation follows that $n$ is bounded:
\begin{equation}
n \leq\frac{1}{2}\Delta
\label{2.37}
\end{equation}
so that only a finite number of terms survive. Using only the singular $\epsilon^{-1}$ terms guarantees that conservation of
the free field currents from which we started is not violated by anomalies. Moreover the conformal covariance of these terms is also maintained. This is the source from which the conformal invariance of the vertex functions follow. There is another consequence: since $n$ is an integer $M+D$ must be even, i.e. $M$ is restricted respectively to even (odd) values if $D$ is even (odd).

The integration over $ q = Q''$
\begin{equation}
\int dq \exp\{-q^2\} \prod_{i=1}^{N} (a_{i}q)^{m_{i}},
\label{2.38}
\end{equation}
is somewhat involved. The trick consists in introducing
\begin{equation}
[\sum_{i=1}^{N} \xi_{i}(a_{i}q)]^{M} = \sum_{m_{i}}{M \choose m_1,m_2,m_3...m_{N}}\prod_{i=1}^{N}\xi_{i}^{m_{i}}
(a_{i}q)^{m_{i}}
\label{2.39}
\end{equation}
and integrating
\begin{eqnarray}
\int dq \exp\{-q^2\} (uq)^{M} = (u^2)^{\frac{1}{2}M} \int d\Omega_{D-1} \cos^{M}\theta \int_{0}^{\infty}d\rho \rho^{M+D-1}
\exp\{-\rho^2\} \nonumber\\
u = \sum_{i=1}^{N}\xi_{i}a_{i}, \quad \mid q \mid = \rho \qquad\qquad
\label{2.40}
\end{eqnarray}
If $M$ is odd, the l.h.s. vanishes since the integrand changes sign under $q\rightarrow -q$. Thus from now on $M$ is assumed to be even, implying also that $D$ must always be even (see (2.35)).
The remaining spherical integral and the integral over $\rho$ are
\begin{eqnarray}
\int d\Omega_{D-1} \cos^{M} \theta = 2 \pi^{\frac{1}{2}(D-1)}\frac{\Gamma(\frac{1}{2}(M+1))}{\Gamma(\frac{1}{2}(M+D))}
\label{2.41} \\
\int_{0}^{\infty} d\rho \rho^{M+D-1}\exp\{-\rho^2\} = \frac{1}{2}\Gamma(\frac{1}{2}(M+D))
\label{2.42}
\end{eqnarray}
From (2.41), (2.42) we recognize that an explicit dependence on $D$ (apart from trivial factors) drops out of our formulae after multiplication of (2.41) with (2.42). We can freely continue our results analytically in $D$, e.g. for $N = 3, s_1=s_2=s_3=4, \Delta_{0} = 4, M_{0} = 8$ we may continue to the value $D=-2$.

Now we expand, neglecting in this first study systematically all the trace terms proportional to $a_{i}^2$ (as was discussed in Section 1) and defining $n_{i,j} = 0$ for $i\geq j$, we get from (2.40)
\begin{equation}
(u^2)^{M/2} = 2^{M/2}\sum_{n_{i,j}}{M/2 \choose \{n_{i,j}\}}\prod_{i<j}(a_{i}a_{j})^{n_{i,j}}\prod_{\ell}\xi_{\ell}^{\sum_{k}(n_{\ell,k}+n_{k,\ell})} +\textnormal{trace terms}
\label{2.43}
\end{equation}
so that by comparison with (2.39)
\begin{equation}
m_{i} =\sum_{k}(n_{i,k} + n_{k,i})
\label{2.44}
\end{equation}
and the coefficient of $\prod_{i,j}(a_{i}a_{j})^{n_{i,j}}$ is
\begin{equation}
2^{M/2}{M/2 \choose \{n_{i,j}\}}{M \choose \{m_{i}\}}^{-1}
\label{2.45}
\end{equation}

The remaining integrations are over $\tau_{i}, i \in \{2,3,...N\}$. They are all elementary. We call $n$ the "order" of the terms. For a first illustration of our algorithm we study mainly the "leading terms" for fixed $n$, this means that we neglect trace and divergence (or deDonder) terms. For arbitray order $n$ the factor $Z^{n}$
yields polynomials of degree $n$ in the differential operators $p_{i}p_{j}$ which for $N=3$ (and only in this case) can be expressed by Laplacians $p_{i}^2$. We emphasize that the leading terms are explicitly independent of $D$, this follows from the multiplication of the two integrals (2.41), (2.42). The traceless and the doubly traceless currents ((2.12)- (2.16)) depend on $D$ explicitly by the parameter $\mu = \frac{D}{2}$.

\section{Evaluating the residues}
\setcounter{equation}{0}
For the leading terms we perform all integrations over the parameters $\tau$.
We start from (2.31), (2.32) where the dependence on the $\tau$ is in the $R_{i}$ and from $Z$, the quadratic complement in the exponent (2.28)
\begin{equation}
Z = \sum_{i=2}^{N}\tau_{i}w_{i-1}^{2}-[\sum_{i=2}^{N}\tau_{i}w_{i-1}]^2
 = \sum_{i,j = 2}^{N} T_{i,j}(w_{i-1},w_{j-1})
\label{3.1}
\end{equation}
where
\begin{equation}
T_{i,j} =\tau_{i}( \delta_{i,j}-\tau_{j})
\label{3.2}
\end{equation}
We use that the integral over the $\tau_{i}$ can be extended to include a variable $\tau_1$ by
\begin{equation}
\int \prod_{i=1}^{N} d\tau_{i}\, \delta(1-\sum_{j=1}^{N} \tau_{j})
\label{3.3}
\end{equation}
which allows us to express $Z$ for $N=3$ in the beautiful symmetric form
\begin{equation}
Z = \tau_1\tau_2 (p_1)^2 + \tau_2\tau_3 (p_2)^2 +\tau_3\tau_1 (p_3)^2
\label{3.4}
\end{equation}
This implies that the terms of order $n$ contribute an $n$'th order polynomial of the Laplacians. For $N>3$ contractions of gradients $(p_{i},p_{j}), i\not=j$, cannot be eliminated.

We emphasize that Euler integrals of the type
\begin{equation}
\int \{\prod_{j=1}^{N} d\tau_{j}\,\tau_{j}^{\alpha_{j}}\,\theta(\tau_{j})\}\,\delta(1-\sum_{i=1}^{N}\tau_{i}) =
\frac{\prod_{j=1}^{N}\alpha_{j}!}{(\sum_{k=1}^{N}\alpha_{k}+N-1)!}
\label{3.5}
\end{equation}
that are easily derived by induction, are the optimal tools for the integration over $\tau_{i}$. Integrals related with these are
\begin{equation}
\int \{\prod _{j=1}^{N} d\tau_{j} \, \tau_{j}^{\alpha_{j}}\,\theta(\tau_{j})\}\,
\theta(1-\sum_{\ell=1}^{N}\tau_{\ell}) = \frac{\prod_{i=1}^{N}\alpha_{i}!}{(\sum_{j=1}^{N}\alpha_{j} +N)!}
\label{3.6}
\end{equation}
Now we turn to the $R_{i}$ terms (2.30)-(2.32). We expand the expression (2.32) in trinomial form
\begin{eqnarray}
(a_{i},Q'+R_{i})^{s_{i}}
= \sum_{k_{i},m_{i}}(-1)^{\eta_{i}}{s_{i} \choose m_{i},k_{i},\eta_{i}}\nonumber\\  \times       (a_{i}Q')^{m_{i}}(a_{i},\frac{1}{2}(w_{i}+w_{i-1}))^{k_{i}} (a_{i}\chi)^{\eta_{i}}\qquad
\label{3.7}
\end{eqnarray}
where
\begin{equation}
\eta_{i} = s_{i}-m_{i}-k_{i}
\label{3.8}
\end{equation}
Besides $Z^{n}$ only the last factor in (3.7) has to be integrated over the $\tau_{i}$.

\subsection{The case $n=0$ for general $N$}
We consider now the simplest case, namely order zero
\begin{equation}
n = 0
\label{3.9}
\end{equation}
Then from the last factor in (3.7) and using (2.29) we get the integral
\begin{equation}
\int (\prod_{j=2}^{N} d\tau_{j})\prod_{i=1}^{N} (a_{i}\chi)^{\eta_{i}} = \int(\prod_{j=2}^{N} d\tau_{j})\prod_{i=1}^{N}\{
\sum_{\{\pi_{i,\alpha}\}}{\eta_{i} \choose \{\pi_{i,\alpha}\}} \prod_{\alpha=1}^{N-1}[\tau_{\alpha+1}\,(a_{i},w_{\alpha})]^{\pi_{i,\alpha}}\}
\label{3.10}
\end{equation}
where the exponents $\pi_{i,\alpha}$ form partitions of $\eta_{i}$, without order and with zeros included.
We introduce the shorthand $\zeta_{\alpha}$ with the properties
\begin{eqnarray}
\sum_{i=1}^{N} \pi_{i,\alpha} = \zeta_{\alpha},\quad
\sum_{\alpha =1}^{N-1} \pi_{i,\alpha} = \eta_{i},\qquad\qquad \nonumber\\ \sum_{\alpha=1}^{N-1} \zeta_{\alpha} = \sum_{i=1}^{N}(s_{i}-m_{i}-k_{i})
= S - M - K, \quad K = \sum_{i} k_{i}
\label{3.11}
\end{eqnarray}
From (3.6) and (3.10) we get by performing the integration
\begin{equation}
\prod_{i=1}^{N}[\sum_{\{\pi_{i,\alpha}\}}{\eta_{i} \choose \{\pi_{i,\alpha}\}}\prod_{\alpha=1}^{N-1}(a_{i}w_{\alpha})^{\pi_{i,\alpha}}]
\frac{\prod_{\beta=1}^{N-1}\zeta_{\beta}!}{[S-M-K +N -1]!}
\label{3.12}
\end{equation}
Multiplying with the trinomial coefficient from (3.7) we can write the coefficients as
\begin{equation}
{s_{i} \choose m_{i},k_{i}, \eta_{i} }{ \eta_{i} \choose \{\pi_{i,\alpha}\}} = {s_{i} \choose m_{i}, k_{i}, \{\pi_{i,\alpha}\}}
\label{3.13}
\end{equation}

Since $M$ is fixed for fixed $n$ (here zero), we obtain contributions where $M/2$ is decomposed into a partition
of numbers $\{n_{i,j}\}$ forming a triangular matrix: $n_{i,j} = 0$ for $i\geq j$ (see (2.43)). For any given set $\{n_{i,j}\}$
the summation parameters $\{m_{i}\}$ (or $\{m_{\ell}\}$) are fixed by (2.44)
\begin{eqnarray}
\Gamma(\frac{M+1}{2})\, 2^{M/2}\sum_{partitions\,\{n_{i,j}\}}{\frac{M}{2} \choose \{n_{i,j}\}}{M \choose \{m_i\}}^{-1}    (\prod_{i,j}(a_{i}a_{j})^{n_{i,j}})\nonumber \\\sum_{partitions \, \{\pi_{\ell,\alpha}\}, k_{\ell}}
\frac{\prod_{\alpha=1}^{N-1}\zeta_{\alpha}!}{(S-M-K +N -1)!}\qquad
\prod_{\ell=1}^{N}(-1)^{\eta_{\ell}}{s_{\ell}\choose m_{\ell},k_{\ell},\{\pi_{\ell,\alpha}\}}\nonumber\\
(\frac{1}{2}(a_{\ell},(w_{\ell}+w_{\ell-1})))^{k_{\ell}}\prod_{\alpha =1}^{N-1}(a_{\ell}w_{\alpha})^{\pi_{\ell,\alpha}}
\qquad\qquad\qquad
\label{3.14}
\end{eqnarray}
The sum over $\alpha$ of the partitions $\pi_{i,\alpha}$ gives by (3.11) the variable $\eta_{i}$. The $\zeta_{\alpha}$
are by (3.11) considered as functions of these partitions in the same way.

The last two factors of (3.14) represent a polynomial of $w$, respectively a differential operator of degree
\begin{equation}
\Delta = S - M, \quad S = \sum_{i}s_{i}
\label{3.15}
\end{equation}

\subsection{The case $n = 1$ for general $N$}
In the case $n=1$ we assume that the number of derivatives is the same, but two of them are reserved now for $Z$.
Therefore we have to replace
\begin{equation}
\Delta = S-M\quad\textnormal{by}\quad \Delta_1 =\Delta -2 =S - M_1 =  S - M - 2
\label{3.16}
\end{equation}
and in (3.14) $M$ by $M_1 = M+2$ everywhere and then insert the factor $Z$
\begin{eqnarray}
\frac{\prod_{\alpha=1}^{N-1} \zeta_{\alpha}!}{(S-M-K+N-1)!}\quad \textnormal{by} \qquad
\frac{\prod_{\alpha=1}^{N-1}\zeta_{\alpha}!}{(S-M-K+N-1)!} \nonumber\\\times \{(S-M-K+N-2)\sum_{\beta=1}^{N-1}(\zeta_{\beta}+1)w_{\beta}^{2} - [\sum_{\beta=1}^{N-1}(\zeta_{\beta}+1)w_{\beta}]^2\}
\label{3.17}
\end{eqnarray}
Instead of (3.14) we obtain then
\begin{eqnarray}
\Gamma(\frac{M+3}{2}) 2^{M/2+1}\sum_{partitions \{n_{i,j}\}}{M/2+1 \choose \{n_{\{i,j\}}\}}{M+2 \choose \{m_{i}\}}^{-1} \prod_{i,j}(a_{i}a_{j})^{n_{i,j}} \sum_{partitions\,\{\pi_{\ell,\alpha}\},k_{\ell}}\nonumber\\
\frac{\prod_{\alpha=1}^{N-1}\zeta_{\alpha}!}{(S-M-K+N-1)!}\{(S-M-K+N-2)\sum_{\beta=1}^{N-1} (\zeta_{\beta}+1)w_{\beta}^{2}\qquad\nonumber\\
-(\sum_{\beta=1}^{N-1}(\zeta_{\beta}+1)w_{\beta})^2\}\prod_{\ell=1}^{N}(-1)^{\eta_{\ell}}{s_{\ell} \choose m_{\ell}, k_{\ell},\{\pi_{\ell,\gamma}\}}\qquad\qquad\nonumber\\ (\frac{1}{2}(a_{\ell},(w_{\ell}+w_{\ell-1})))^{k_{\ell}}\prod_{\gamma=1}^{N-1}(a_{\ell}w_{\gamma})^{\pi_{\ell,\gamma}}
\qquad\qquad\qquad
\label{3.18}
\end{eqnarray}
In general $n$ extends from zero to $\frac{\Delta}{2}$ (see (1.9), (2.37)).

\section{The free Lagrangian and the 2-vertex}
\setcounter{equation}{0}
The general treatment of the Noether potential for arbitrary $N$ in Section 3 can be applied to the case
$N=2$, too. In this case the two higher spin fields must be identical by QFT. It sounds reasonable to use the double traceless currents (2.12),(2.14)-(2.16) including also the B-parameter, to check the trace terms of the 2-vertex against the trace terms of the free Lagrangian. We obtain immediately
\begin{eqnarray}
<0\mid \tilde J^{(s)}(x_1;a_1) \tilde J^{(s)}(x_2;a_2) \mid 0> \nonumber\\
= \sum_{n=0}^{s/2} \alpha_{n}^{(s)2} (a_1^2 \, a_2^2)^{n}\,<0\mid J_{n}^{(s-2n)}(x_1;a_1) J_{n}^{(s-2n)}(x_2;a_2)\mid 0>\nonumber\\
= \{\sum_{n=0}^{s/2} [\frac{\alpha_{n}^{(s)}}{[s]_{2n}}]^2\, (a_1^2 a_2^2)^{n}(\Box_{a_1}\Box_{a_2})^{n}\} <0\mid j^{(s)}(x_1;a_1) j^{(s)}(x_2;a_2) \mid0>
\label{4.1}
\end{eqnarray}
So we have obtained one differential operator in the tensorial spaces of $a_1,a_2$ applied to one vacuum expectation value.
If this 2-point function is contracted with two equal higher spin fields of spin $s$, which are double traceless, then in
(4.1) all terms in the differential operator containing powers $(a_1^2a_2^2)^{n},\, n\geq 2$ drop out. On the other hand
$\alpha_1^{(s)}$ vanishes by (2.14). Therefore the whole differential operator in (4.1) reduces to the unit
operator. If we allow next for the B-parameter, we use (2.18) to resume it in the higher spin field (see below).

Since $\Delta = 2$ admits only the two poles at $n = 0$ and $n = 1$, we can simply carry over the results of Section 3. The main generalization we must admit concerns the matrix ${n_{i,j}}$ of exponents of $(a_{i},a_{j})$ which is upper triangular still but now with a nonvanishing diagonal for the traces, for $N=2$
\begin{eqnarray}
n_{i,j} = 0 \quad\textnormal{if}\quad i>j \qquad\qquad \nonumber\\
m_1 = 2n_{1,1} + n_{1,2}, \quad m_2 = 2n_{2,2}+n_{1,2}
\label{4.2}
\end{eqnarray}
But we have to modify the eqs (3.14), (3.18) in the powers of $2$: Replace in (3.14) $2^{M/2}$ and in (3.18) $2^{M/2+1}$ by         the same expression $2^{\sum_{i<j}n_{i,j}}$.

Then in (3.14) we consider the last two factors. There is only one momentum variable contained in them
\begin{equation}
w_1 = p_1 = -p_2 = p
\label{4.3}
\end{equation}
and these factors take the form
\begin{equation}
2^{-(k_1+k_2)}(a_1,p)^{k_1+\pi_{1,1}}(a_2,p)^{k_2+\pi_{2,1}} = 2^{-(k_1+k_2)}(a_1,p)^{s-m_1}(a_2,p)^{s-m_2}
\label{4.4}
\end{equation}

\subsection{The case $n=0$}
From $\Delta = 2$ follows $M = 2s -\Delta = 2s -2$. For the exponents $\{n_{i,j}\}$ the following list of
possible values exists
\begin{eqnarray}
\begin{array}{cccccccc}
   & n_{1,1} &  n_{2,2} &  n_{1,2} &  m_1  &   m_2 & k_1+\pi_{1,1} & k_2+\pi_{2,1}   \\
(1)&    0  &     0      &   s-1    &  s-1  &   s-1   &     1   &        1            \\
(2)&    1  &  0         &   s-2    &  s  &   s-2     &     0   &        2            \\
(3)&    0  &  1         &   s-2      &  s-2&   s     &     2   &        0            \\
(4)&    1  &  1         &   s-3    &  s-1  &  s-1    &     1   &        1            \\
\end{array}
\label{4.5}
\end{eqnarray}
For each of these four cases (1)-(4) we have the following possible values of $k_1,\pi_{1,1},k_2, \pi_{2,1}$
\begin{eqnarray}
\begin{array}{ccccc}
     &  k_1  &  k_2  &  \pi_{1,1}  &  \pi_{2,1}  \\
(1),(4)  &   1   &   1   &    0        &      0      \\
(1),(4)  &   1   &   0   &    0        &      1      \\
(1),(4)  &   0  &  1     &    1        &      0      \\
(1),(4)  &   0  &  0     &    1        &      1      \\
(2)  &   0  &  2     &    0        &      0      \\
(2)  &   0  &  1     &    0        &      1      \\
(2)  &   0  &  0     &    0        &      2      \\
(3)  &   2  &  0     &    0        &      0      \\
(3)  &   1  &  0     &    1        &      0      \\
(3)  &   0  &  0     &    2        &      0      \\
\end{array}
\label{4.6}
\end{eqnarray}
These we have to insert into the modified formula (3.14)
\begin{eqnarray}
\Gamma(\frac{1}{2} (2s-1)) \sum_{n_{1,1},n_{1,2}, n_{2,2}, k_1, k_2,\pi_{1,1}, \pi_{2,1}} 2^{n_{1,2} -k_1-k_2}
{s-1 \choose n_{1,1},n_{1,2},n_{2,2}}{2s-2 \choose m_1,m_2}^{-1} \nonumber\\ \times (-1)^{\pi_{1,1} + \pi_{2,1}}\frac{(\pi_{1,1} +\pi_{2,1})!} {(3-k_1-k_2)!}{s \choose m_1,k_1,\pi_{1,1}}{s \choose m_2,k_2,\pi_{2,1}}\nonumber\\
\times (a_1^2)^{n_{1,1}}(a_2^2)^{n_{2,2}}(a_1,a_2)^{n_{1,2}}\,(a_1,p)^{s-m_1}(a_2,p)^{s-m_2} \qquad\qquad
\label{4.7}
\end{eqnarray}
The result consists of three terms
\begin{eqnarray}
\frac{s!\sqrt{\pi}}{4!2^{s}} \{4s(a_1,p)(a_2,p)(a_1,a_2)^{s-1}\qquad\qquad\qquad \nonumber\\
+s(s-1) [ (a_2,p)^2\, a_1^2 + (a_1,p)^2\,a_2^2]\,(a_1,a_2)^{s-2} \nonumber\\
+s(s-1)(s-2)(a_1,p) \,(a_2,p)\, (a_1,a_2)^{s-3} \, a_1^2\,a_2^2 \} \qquad
\label{4.8}
\end{eqnarray}

\subsection{The case $n=1$}
The derivatives are now all in
\begin{equation}
(w_1)^2 = p^2
\label{4.9}
\end{equation}
so that
\begin{equation}
k_1+\pi_{1,1} = k_2 + \pi_{2,1} = 0
\label{4.10}
\end{equation}
It follows that only two possibilities appear
\begin{eqnarray}
\begin{array}{ccccc}
         & n_{1,1} & n_{2,2} & n_{1,2} & {s \choose n_{1,1},\,n_{2,2},\,n_{1,2} } \\
(\alpha)  &    0    &   0     &    s    &           1    \\
(\beta)   &    1    &   1     &    s-2  &           s(s-1) \\
\label{4.11}
\end{array}
\end{eqnarray}
which yield the expression
\begin{equation}
-\frac{s!\,\sqrt{\pi}}{4!\,2^{s}} \, p^2 \, [4(a_1,a_2)^{s} +s(s-1) (a_1,a_2)^{s-2} \, a_1^2 \, a_2^{2}]
\label{4.12}
\end{equation}
Thus we obtain for the sum of the terms (4.8) and (4.12)
\begin{eqnarray}
\frac{s! \sqrt{\pi}}{4! 2^{s}}\{ [4s(a_1,a_2)^{s-1} +s(s-1)(s-2)(a_1,a_2)^{s-3}a_1^2 a_2^2](a_1,p)(a_2,p)\nonumber\\
 +s(s-1) [(a_1,p)^2 a_2^2 +(a_2,p)^2 a_1^2]\,(a_1,a_2)^{s-2} \qquad\qquad\nonumber\\-p^2 [4(a_1,a_2)^{s} +s(s-1) (a_1,a_2)^{s-2}\,a_1^2a_2^2 ]\}\qquad\qquad
\label{4.13}
\end{eqnarray}
The free Lagrangian is on the other hand
\begin{eqnarray}
\mathcal{L}_0 [h^{(s)}] = -\frac{1}{2} h^{(s)}(x;a) \ast_{a} \mathcal{F}^{(s)}(x;a) \nonumber\\
+\frac{1}{8s(s-1)}\Box_{a}h^{(s)}(x;a) \ast_{a}\Box_{a} \mathcal{F}^{(s)}(x;a)
\label{4.14}
\end{eqnarray}
where the Fronsdal term is defined as
\begin{eqnarray}
\mathcal{F}^{(s)}(x;a) = \Box h^{(s)}(x;a) -(a\nabla)(\nabla\partial_{a})h^{(s)}(x;a) +\frac{1}{2}(a\nabla)^2 \Box_{a} h^{(s)}(x;a)\\
\label{4.15}
\Box_{a} \mathcal{F}^{(s)}(x;a) = 2 [\Box \Box_{a} -(\nabla\partial_{a})^2]h^{(s)}(x;a)\qquad\qquad
\label{4.16}
\end{eqnarray}
In (4.13) we must replace $a_{i}$ by $\partial_{a_{i}}$ and apply the resulting differential
operator to $h^{(s)}(x;a_1)h^{(s)}(x;a_2)$.

First we notice that the free parameter $B_1^{(s)}$ enhances the problem with the trace terms if it does not vanish. Namely with the trace field $\bar h^{(s-2)}= [s(s-1)]^{-1}\Box_{a}h^{(s)}$ for a doubly traceless field $h^{(s)}$ we get
\begin{equation}
\Box_{a}[ h^{(s)} + \frac{B_1^{(s)}}{s(s-1)}a^2\Box_{a}h^{(s)}] = [1+ \frac{2B_1^{(s)}(D+2s-4)}{s(s-1)}]\Box_{a} h^{(s)}
\label{4.17}
\end{equation}
and we obtain a coefficient depending on the space dimension $D$. Such coefficients are unknown for $N=2$ or $N=3$.

However, if we neglect the trace terms $\bar h^{(s-2)}$, the Fronsdal field (4.15) reduces to
\begin{equation}
\mathcal{F}^{(s)}(x;a)\mid _{\bar h^{(s-2)} = 0} = \Box h^{(s)}(x;a) - (a\nabla)(\nabla\partial_{a}) h^{(s)}(x;a)
\label{4.18}
\end{equation}
and (4.13) simplifies to
\begin{equation}
\frac{(s!)^3 \sqrt \pi}{ 6\cdot 2^{s}} h^{(s)}(x;a) \ast_{a} \mathcal{F}(x;a)\mid_{\bar h^{(s)} = 0}
\label{4.19}
\end{equation}
which is, up to the normalization, the same as (4.14).

\section{The cubic Noetherian}
\subsection{The leading terms}
\setcounter{equation}{0}
In \cite{MMR2} the leading terms of the cubic Lagrangian (those which remain when the trace and deDonder fields vanish)
are presented in the cyclic form
\begin{eqnarray}
\mathcal{L}^{(0,0)} _{N}[h] = \sum_{n_{i}} C^{s_1,s_2,s_3}_{n_1,n_2,n_3}\int dz_1dz_2dz_3 \delta(z_3-z_1)\delta(z_2-z_1)\nonumber\\
\times(\partial_{a}\partial_{b})^{Q_{1,2}}(\partial_{b}\partial_{c})^{Q_{2,3}}(\partial_{c}\partial_{a})^{Q_{3,1}}(\partial_{a}\nabla_2)^{n_1}(\partial_{b}\nabla_3)^{n_2} \nonumber\\ \times(\partial_{c}\nabla_1)^{n_3} h^{(s_1)}(z_1;a) h^{(s_2)}(z_2;b) h^{(s_3)}(z_3;c)\qquad
\label{5.1}
\end{eqnarray}
The total order of space derivatives is
\begin{equation}
\Delta = n_1+n_2+n_3
\label{5.2}
\end{equation}
and the equations for the tensorial orders are
\begin{equation}
n_1+Q_{1,2} +Q_{3,1} = s_1, \,
n_2+Q_{2,3} +Q_{1,2} = s_2, \,
n_3+Q_{3,1} +Q_{2,3} = s_3
\label{5.3}
\end{equation}
These equations are solved by
\begin{equation}
Q_{2,3}= n_1-\nu_1,\, Q_{3,1}= n_2-\nu_2, \, Q_{1,2}= n_3-\nu_3, \, n_{i} \geq \nu_{i} \geq 0
\label{5.4}
\end{equation}
implying
\begin{equation}
\nu_{i} = s_{i} -\frac{1}{2}(S-\Delta),\quad S = \sum_{i=1}^3\, s_{i}
\label{5.5}
\end{equation}
Since from \cite{Met} we know that the minimal order $\Delta$ is
\begin{equation}
\Delta_{min} = S -2s_3 \quad \textnormal{in the case} \quad s_1\geq s_2 \geq s_3
\label{5.6}
\end{equation}
we obtain that in this case
\begin{equation}
\nu_1 = s_1-s_3,\, \nu_2 = s_2-s_3,\, \nu_3 = 0
\label{5.7}
\end{equation}
From now on we consider only the case of minimal $\Delta_{min}$ (5.6).
Finally we give the coefficient
\begin{equation}
C^{s_1,s_2,s_3}_{n_1,n_2,n_3} = {\sum_{i,j}  Q_{i,j} \choose Q_{1,2}, Q_{2,3}, Q_{3,1} },\quad \sum_{i,j}Q_{i,j} =\frac{S-\Delta_{min}}{2} = \frac{M}{2} = s_3
\label{5.8}
\end{equation}

\subsubsection{The case n=0}
We concentrate for simplicity on the example $\Delta_{min}$.
In the cubic case the exponents $m_{i}$ and $n_{i,j}$ form triplets and these can be mapped one-to-one onto each other
and on the variables $Q_{i,j}$ and $n_{i}$. These relations are
\begin{eqnarray}
n_{12} = Q_{12} = n_3 - \nu_3 = s_3-m_3  \nonumber\\
n_{13} = Q_{31} = n_2 - \nu_2 = s_3-m_2\nonumber\\
n_{23} = Q_{23} = n_1 - \nu_1 = s_3-m_1 \nonumber\\
\nu_{i} = \frac{1}{2}(\Delta +s_{i}-s_{j}-s_{k}),\, i,j,k \,\textnormal{different}
\label{5.9}
\end{eqnarray}
so that the trinomial coefficient in (5.8) agrees with the trinomial factor in (2.43).
These relations represent the first step towards a reconstruction of the structure of the cubic Lagrangian.

Now we consider the $w$-polynomials contained in (3.14), (3.18) leading to the differential operators. We neglect all divergence expressions
of the type
\begin{equation}
(a_{i}p_{i}) = 0
\label{5.10}
\end{equation}
together with all trace terms proportional to any $a^{2}$
\begin{equation}
\frac{1}{2} (a_{\ell}, w_{\ell} +w_{\ell-1}) = (a_{\ell}, w_{\ell-1})
\label{5.11}
\end{equation}
and in detail we get by inserting $w_{i}$ from (2.9)
\begin{eqnarray}
(a_1,w_1) =  0, \quad (a_2,w_1) = -(a_2,p_3),\quad (a_3,w_1) = (a_3,p_1) \nonumber\\
(a_1,w_2) =  (a_1,p_2), \quad (a_2,w_2) = -(a_2, p_3), \quad (a_3,w_2) = 0 \nonumber\\
\qquad
\label{5.12}
\end{eqnarray}
Consequently in (3.14) we obtain
\begin{equation}
k_1 = k_3 = 0, \quad \pi_{1,1} = \pi_{3,2} = 0
\label{5.13}
\end{equation}
implying
\begin{equation}
\pi_{1,2} = \eta_1 = s_1-m_1, \quad \pi_{2,1} + \pi_{2,2} = \eta_2= s_2-m_2-k_2, \quad \pi_{3,1} = \eta_3 = s_3-m_3
\label{5.14}
\end{equation}

Thus the $w$-polynomial comes out as
\begin{equation}
(a_1,p_2)^{s_1-m_1}(a_2,p_3)^{s_2-m_2}(a_3,p_1)^{s_3-m_3}
\label{5.15}
\end{equation}
which leads to the differential operator expected from the balance equations and the cyclic ordering.
Thus there remains only to check the numerical factor. The coefficients in front are
\begin{equation}
{s_3 \choose n_{1,2},n_{2,3},n_{1,3} }{2s_3 \choose m_1,m_2,m_3 }^{-1}
\label{5.16}
\end{equation}
The desired final trinomial coefficient is the first factor of (5.16), which was stated already after (5.9), meaning that the second trinomial coefficient must be cancelled by the other coefficients in (3.14). We will see that to achieve this, two summations have to be done in analytic terms.

The further factors that we find for fixed $m_1, m_2, m_3$ depend on
\begin{equation}
\zeta_1 = \pi_{2,1}+\pi_{3,1} = \sigma_{2,3}-k_2 -\pi_{2,2}, \quad \zeta_2 = s_1-m_1 +\pi_{2,2}, \quad \sigma_{i,j} = (s_{i}-m_{i}) +(s_{j}-m_{j})
\label{5.17}
\end{equation}
and we have to evaluate (the replacement  $(a_2,p_1) =-(a_2,p_3)$ contributes to the phase factor)
\begin{eqnarray}
\sum_{k_2,\pi_{2,2}}(-1)^{m_2+k_2}{s_1\choose m_1}{s_2 \choose m_2}{s_3 \choose m_3}{s_2-m_2 \choose k_2,\pi_{2,2}, s_2-m_2-k_2-\pi_{2,2}}\nonumber\\
\frac{(s_1-m_1+\pi_{2,2})!(\sigma_{2,3}-k_2-\pi_{2,2})!}{(\Delta+2-k_2)!}(a_1,p_2)^{s_1-m_1} (a_2,p_3)^{s_2-m_2}(a_3,p_1)^{s_3-m_3}
\label{5.18}
\end{eqnarray}
where we must sum over $k_2$ and $\pi_{2,2}$ finally. The result is, after evaluating two finite $_2F_1(1)$ series and applying the well known identity for $\Gamma(z)\Gamma(1-z)$
\begin{equation}
\frac{1}{(\Delta+2)!} (s_1-m_1)!(s_2-m_2)!(s_3-m_3)!{s_1 \choose m_1}{s_2\choose m_2}{s_3 \choose m_3}\prod_{i}(a_{i},p_{i+1})^{s_{i}-m_{i}}
\label{5.19}
\end{equation}

Finally we take the remaining factors into account and cancel the three binomials in (5.11) after multiplication with the second factor of (5.16)
with the result
\begin{equation}
\frac{s_1! s_2!s_3!}{(2s_3)!(\Delta+2)!}{s_3\choose n_{1,2},n_{2,3},n_{1,3}}
\label{5.20}
\end{equation}
which is the result expected from \cite{MMR2} up to the normalizing factor and coincides with (5.8).

\subsubsection{The cases $n\geq1$}
For $N=3$ we use (3.4) and expand $Z^{n}$ where $2n$ is bounded by $\Delta_0$
\begin{equation}
Z^{n} = \sum_{\lambda_{i}}{n \choose \lambda_1,\lambda_2,\lambda_3}\tau_1^{\lambda_1+\lambda_3}\tau_2^{\lambda_1+\lambda_2}
\tau_3^{\lambda_2+\lambda_3}
(p_1^{2})^{\lambda_1}(p_2^{2})^{\lambda_2}  (p_3^{2})^{\lambda_3}
\label{5.21}
\end{equation}
Instead of (3.18) we obtain now after integration over the $\tau_{i}$ and inserting $M_{n} = M_{0} +2n, \Delta_{n} = \Delta_0 -2n,
M_0 = S -\Delta_0 = 2s_3$ (we use primed quantities for $m_{i}$ and $n_{i,j}$ to mark the dependence on $n$)
\begin{eqnarray}
\Gamma(\frac{M_0+1}{2}+n)\,2^{\frac{M_0}{2}+n}\,{\frac{M_0}{2}+n \choose \{n_{i,j}'\}}{M_0+2n\choose \{m_{i}'\}}^{-1} \prod_{i,j}
(a_{i},a_{j})^{n_{i,j}'}\sum_{partitions \,\{\pi_{\ell,\alpha}\},\, k_{\ell}}\qquad\nonumber\\
\frac{(-1)^{n}}{n!}\sum_{\lambda_{i}}{n \choose \lambda_1,\lambda_2,\lambda_3} \frac{(\lambda_1+\lambda_3)!(\lambda_1+\lambda_2+\zeta_1)!(\lambda_2+\lambda_3+\zeta_2)!}{(S-M_0 -K +2)!}\qquad\qquad\nonumber\\
(p_1^{2})^{\lambda_1}(p_2^{2})^{\lambda_2} (p_3^{2})^{\lambda_3}\, \prod_{\ell=1}^{3}(-1)^{\eta_{\ell}}{s_{\ell}\choose m_{\ell}',k_{\ell},\{\pi_{\ell,\gamma}\}} \qquad\qquad\qquad\nonumber\\(\frac{1}{2}(a_{\ell},(w_{\ell}+w_{\ell-1})))^{k_{\ell}}\prod_{\gamma = 1}^{2}
(a_{\ell},w_{\gamma})^{\pi_{\ell,\gamma}} \qquad\qquad\qquad \label{5.15}
\end{eqnarray}
following the same arguments (5.3)-(5.5) as in the case $n=0$.
The group of terms we are considering is
\begin{eqnarray}
[\prod_{\ell =1}^{3}(-1)^{\eta_{\ell}} {s_{\ell} \choose m_{\ell}',k_{\ell}, \{\pi_{\ell,\alpha}\}}]
(a_1,p_2)^{\pi_{1,2}} (a_2,p_3)^{\pi_{2,1}+\pi_{2,2} +k_2}(a_3,p_1)^{\pi_{3,1}}\nonumber\\ = (-1)^{m_2'+k_2}[\prod_{i=1}^{3}{s_{i}\choose m_{i}'}]{s_2-m_2' \choose k_2, \pi_{21}, \pi_{22}}(a_1,p_2)^{\pi_{1,2}}(a_2,p_3)^{\pi_{2,1}+\pi_{2,2}+k_2}(a_3,p_1)^{\pi_{3,1}}
\label{5.23}
\end{eqnarray}
Thus it remains to do the summations over $k_2$ and $\pi_{2,2}$. Using the results (5.3) - (5.7) we get first the
polynomial in $w_{i}$ (5.8) which is now of degree $\Delta_{n}$. The differential operator that is only a polynomial in the Laplacians results from the sum
\begin{eqnarray}
(-1)^{n}\sum_{\pi_{2,2},k_2}\,[(S-M_0-k_2+2)!]^{-1} \,\sum_{\{\lambda_{i}\}, \sum_{i}\lambda_{i} = n}\frac{(p_1^{2})^{\lambda_1}}{\lambda_1!}\frac{(p_2^2)^{\lambda_2}}{\lambda_2!}
\frac{(p_3^{2})^{\lambda_3}}{\lambda_3!} \nonumber\\
(\lambda_1+\lambda_3)!(\lambda_1+\lambda_2+\zeta_1)!(\lambda_2+\lambda_3+\zeta_2)! \label{4.17}\qquad\qquad\qquad
\label{5.24}
\end{eqnarray}
where (as in (5.17))
\begin{equation}
\zeta_1 = \sigma_{2,3}-k_2-\pi_{2,2},\, \zeta_2 = s_1-m_1'+\pi_{2,2},\, \sigma_{i,j} = s_{i}-m_{i}' +s_{j}-m_{j}'
\label{5.25}
\end{equation}
The technique of summation is the same as in the case $n=0$ and we get altogether
\begin{eqnarray}
(-1)^{n}\Gamma(s_3+n +1/2)\, 2^{s_3+n}{s_3+n \choose \{n_{i,j}'\}}{2s_3+2n \choose \{m_{i}'\}}^{-1}\, \{\prod_{k=1}^{3}{s_{k}\choose m_{k}'}\}[(\Delta_0+2)!]^{-1}\nonumber\\\sum_{\lambda_{i},\sum_{i}\lambda_{i}=n}(s_1-m_1'+\lambda_2+\lambda_3)!(s_2-m_2' +\lambda_1+\lambda_3)! (s_3-m_3'+\lambda_1+\lambda_2)!\nonumber\\ \{\prod_{r,q}(a_{r},a_{q})^{n_{r,q}'}\}(a_1,p_2)^{s_1-m_1'}(a_2,p_3)^{s_2-m_2'} (a_3,p_1)^{s_3-m_3'} \prod_{\ell=1}^{3}\frac{(p_{\ell}^{2})^{\lambda_{\ell}}}{
\lambda_{\ell}!}\qquad\qquad \label{5.26}
\end{eqnarray}
A certain simplification of this expression is possible by
\begin{eqnarray}
{2s_3+2n \choose m_1',m_2',m_3'}^{-1}\{\prod_{k=1}^{3}{s_{k}\choose m_{k}'}\}(s_1-m_1'+\lambda_2+\lambda_3)!\qquad\qquad\nonumber\\ \times(s_2-m_2'+\lambda_1+\lambda_3)!(s_3-m_3'+\lambda_1+\lambda_2)! =\{(2s_3+2n)!\}^{-1}
s_1!s_2!s_3!\nonumber\\ \times
(s_1-m_1'+1)_{\lambda_2+\lambda_3} (s_2-m_2'+1)_{\lambda_1+\lambda_3}(s_3-m_3'+1)_{\lambda_1+\lambda_2}\qquad\qquad
\label{5.27}
\end{eqnarray}
The sum over $n$ of (5.26) vanishes, as long as divergences and traces are neglected, on shell and belongs to the kernel of the Noetherian, or can be eliminated by a field redefinition. The situation changes if we admit divergence terms.

\subsection{The divergence terms}
We allow for divergence terms
\begin{equation}
(p\partial_{a})\hat h^{(s)}(p;a)
\label{5.28}
\end{equation}
To extract trace field terms instead of divergence field terms would be more involved, since these arise from both the proper trace field and the deDonder field terms. In turn the deDonder field terms can be obtained by implementing a trace term into each divergence term as defined in (1.1), so that one trace field term results from each divergence term. But there are also the proper trace field terms. However, the divergence terms have also two different sources.

\subsubsection{The first source}
First we consider the group of terms from $n=0$ treated in subsection 5.1 and make the divergence terms explicit that were neglected there. In \cite{MMR2} the group of terms left over when we annihilate all traces, form the four blocks
\begin{equation}
\mathcal{L}^{(0,0)},\quad \mathcal{L}^{(1,0)},\quad \mathcal{L}^{(2,0)}, \quad \mathcal{L}^{(3,0)}
\label{5.29}
\end{equation}
Then equs. (2.2) and (3.16)-(3.18) of \cite{MMR2} can be reformulated as
\begin{eqnarray}
 \{\sum_{i=0}^3\mathcal{L}^{(i,0)}\mid_{\textnormal{traces} = 0}\}\qquad\qquad\qquad \nonumber\\ =[1 +\frac{1}{2} \{\sum_{i} (a_{i},p_{i})\frac{\partial}{\partial(a_{i},p_{i+1})}\qquad\qquad\qquad \nonumber\\
+\sum_{i<j}(a_{i},p_{i})(a_{j},p_{j})\frac{\partial^{2}}{\partial(a_{i},p_{i+1})\partial(a_{j},p_{j+1})}\qquad\qquad\nonumber\\
+ (a_1,p_1)(a_2,p_2)(a_3,p_3) \frac{\partial^{3}}{\partial(a_1,p_2)\partial(a_2,p_3) \partial(a_3,p_1)}\}] \mathcal{L}^{(0,0)}
\label{5.30}
\end{eqnarray}
Comparison with (5.1) the curly bracket in (5.30) yields the differential operator part
\begin{equation}
\frac{1}{2} \sum_{\delta_1,\delta_2,\delta_3}\prod_{i=1}^3[n_{i}(a_{i},p_{i})]^{\delta_{i}}(a_{i},p_{i+1})^{n_{i}-\delta_{i}}
\quad (p_{i+1}\mid_{i=3} = p_1),\quad n_{i} = s_{i} -m_{i}
\label{5.31}
\end{equation}
where $\delta_{i}$ is either zero or one and $\sum_{i=1}^3 \delta_{i}\geq 1$.

To obtain these divergence terms from the $n=0$ expression in subsection 5.1.1 (see (2.22),(5.23)) we expand, keeping divergence terms
\begin{eqnarray}
\sum_{\kappa_2,\lambda_{1,2},\lambda_{2,1}} 2^{-(k_1+\kappa_2+k_3)}(-1)^{k_2+k_3+\pi_{2,1}+\pi_{2,2}+\pi_{3,2}}
{k_2\choose\kappa_2}{\pi_{2,1} \choose \lambda_{2,1}}{\pi_{1,2} \choose \lambda_{1,2}}\nonumber\\
(a_1,p_1)^{k_1+\pi_{1,1}+\lambda_{1,2}} (a_2,p_2)^{\lambda_{2,1}+\kappa_2} (a_3,p_3)^{\pi_{3,2}+k_3}\nonumber\\
(a_2,p_3)^{\pi_{2,1}-\lambda_{2,1}+\pi_{2,2}+k_2-\kappa_2}(a_3,p_1)^{\pi_{3,1}}(a_1,p_2)^{\pi_{1,2}-\lambda_{1,2}}
\label{5.32}
\end{eqnarray}
We denote the powers of the divergence terms by
\begin{equation}
\delta_1 = k_1+\pi_{1,1}+\lambda_{1,2},\, \delta_2 =  \kappa_2 + \lambda_{2,1}, \, \delta_3 = k_3 +\pi_{3,2}
\label{5.33}
\end{equation}
Then the last line of (5.32) can, see (5.14), (5.15)), be written as
\begin{equation}
(a_2,p_3)^{s_2-m_2-\delta_2} (a_3,p_1)^{s_3-m_3-\delta_3} (a_1,p_2)^{s_1-m_1-\delta_1}
\label{5.34}
\end{equation}
which agrees with (5.31).

The operator form being correct, we set out to calculate the coefficients. We will do this here only for the case
\begin{equation}
\delta_1 =1,\, \delta_2 = \delta_3 = 0.
\label{5.35}
\end{equation}
If this coefficient comes out as predicted by (5.30), (5.31), we have no doubt that the coefficients in all other cases come out correctly, too.

We proceed as follows. From (5.33) we have only three choices for the parameters $k_1, \pi_{1,1}$
\begin{equation}
k_1 \in \{0,1\}, \, \pi_{1,1} \in \{0,1\},\, k_1 \cdot \pi_{1,1} = 0
\label{5.36}
\end{equation}
We treat these cases separately. Then in each case we have to do $_2F_1(1)$ summations in the way we did before. Let us
formulate the program by
\begin{eqnarray}
(s_1-m_1)\sum_{k_1,k_2,k_3,\pi_{1,1},\pi_{2,2},\pi_{3,2}}\qquad\qquad\qquad\nonumber\\
\{\delta_{k_1,0}\delta_{\pi_{1,1},0}\frac{(\sigma_{2,3}-k_2-\pi_{2,2})!(s_1-m_1+\pi_{2,2})!}{(\Delta_{0}+2-k_2)!}\qquad\nonumber\\
+\delta_{k_1,0}\delta_{\pi_{1,1},1} \frac{(\sigma_{2,3}-k_2-\pi_{2,2}+1)!(s_1-m_1+\pi_{2,2}-1)!}{(\Delta_{0} +2 -k_2)!}\nonumber\\
-\frac{1}{2} \delta_{k_1,1}\delta_{\pi_{1,1},0}\frac{(\sigma_{2,3}-k_2-\pi_{2,2})!(s_1-m_1+\pi_{2,2}-1)!}{(\Delta_{0} +1 -k_2)!}\}
\nonumber\\ \times (-1)^{m_2+k_2}{s_2-m_2 \choose k_2,\pi_{2,1},\pi_{2,2}} \delta_{k_3,0}\delta_{\pi_{3,2},0}
\label{5.37}
\end{eqnarray}
with $\sigma_{i,j}$ from (5.17). The summations are performed in the same fashion as in subsection 5.1 and the result is
\begin{eqnarray}
\frac{1}{(2s_3)! (\Delta_{0}+2)!}{s_3 \choose n_{1,2},n_{1,3},n_{2,3}}\prod_{i=1}^{3} (s_{i}-m_{i})! \nonumber\\
\times\frac{1}{2} \{(s_1-m_1) -(s_2-m_2) + (s_3-m_3)\}
\label{5.38}
\end{eqnarray}
where the factorials cancel against factorials from
\begin{equation}
\prod_{i=1}^{3} {s_{i} \choose m_{i}}
\label{5.39}
\end{equation}
Comparing this result with (5.20) we obtain as relevant factor the second line of (5.38).

\subsubsection{The second source}
This group of terms arises from $n\geq 1$. For an evaluation it is more practical to use
\begin{equation}
(\partial_{a_{i}},\partial_{a_{j}}) \, (\partial_{a_{i}}, p_{j})
\label{5.40}
\end{equation}
acting on the fields $h^{(s)}(x_{i};a_{i})$ as differential operators. We start from Section 5.1.1 and our
aim is to replace the Laplacians $p_{i}^2$ by the tracefree Fronsdal fields
\begin{equation}
\mathcal{F}_{i} = [-p_{i}^2 +(a_{i},p_{i})(p_{i},\partial_{a_{i}})]\hat h^{(s)}(p;a)
\label{5.41}
\end{equation}
In this fashion we must, after neglecting all Fronsdal fields (e.g. eliminating them by a field redefinition), replace the last factor in (5.26) by
\begin{equation}
\mathcal{G}(\Lambda) = \prod_{\ell = 1}^{3} \frac{[(a_{\ell},p_{\ell})(p_{\ell},\partial_{a_{\ell}})]^{\lambda_{\ell}}}{\lambda_{\ell}!},\quad \Lambda = \{\lambda_1,\lambda_2,\lambda_3\}
\label{5.42}
\end{equation}
Because there is at most one divergence operation of each type (two of the same type vanish on shell and are, similarly as the Fronsdal operators, not resumed in the ansatz for a Lagrangian), of interest are only the cases
\begin{equation}
\lambda_{i} \leq 1
\label{5.43}
\end{equation}
selected from pole terms $n$ up to three.

Now we consider the case $\Lambda = (1,0,0)$ which in (5.26) necessitates only to analyse the case $n=1$.
If we suppose that all Fronsdal operators are directly acting on the field $h$, it follows that the operators
$(a_{i},p_{i})(p_{i},\partial_{a_{i}})$ arising act also directly on these fields. This implies that the gradient
operator, namely the first factor in this combination, stands to the right of all differential operators that appear in the same order as in (5.26)
\begin{equation}
\prod_{r,q} (\partial_{a_{r}},\partial_{a_{q}})^{n_{r,q}'}\,\prod_{i=1}^{3} (\partial_{a_{i}} p_{i+1})^{s_{i}-m_{i}'}
\label{5.44}
\end{equation}
Our strategy is then to move it to the left by commutators. At the left end we get zero because we get $s+1$ derivations with respect to $a$ on $h^{(s)}$. A commutator of the gradient $(a_1,p_1)$ with $(\partial_{a_{1}},p_{2})$ gives the sum of three Laplacians which lead to Fronsdal and further divergence fields. They can be neglected here. Thus we have only to deal with the commutator
\begin{eqnarray}
[\prod_{i,j}(\partial_{a_{r}},\partial_{a_{q}})^{n_{i,j}'}, (a_1,p_1)] = \qquad\qquad\qquad\nonumber\\
(\partial_{a_2},\partial_{a_3})^{n_{2,3}'}\{n_{1,2}'(\partial_{a_1},\partial_{a_3})^{n_{1,3}'}(\partial_{a_1},\partial_{a_2})^{n_{1,2}'-1}(\partial_{a_2},p_1) + \nonumber\\
n_{1,3}' (\partial_{a_1},\partial_{a_3})^{n_{1,3}'-1}(\partial_{a_1},\partial_{a_2})^{n_{1,2}'}(\partial_{a_3},p_1)\}\qquad
\qquad
\label{5.45}
\end{eqnarray}
Neglecting once again an additional divergence operator we replace $(\partial_{a_2},p_1)$ by $-(\partial_{a_2},p_3)$.

The next task is to identify the parameters $\{n_{i,j}', m_{i}'\}$ by the $\{n_{i,j},m_{i}\}$. To do this we have to treat the two cases included in (5.45) separately. The term (I) (the second in (5.45)) is
\begin{eqnarray}
(I):\qquad\qquad +n_{1,3}'(\partial_{a_1},\partial_{a_2})^{n_{1,2}'}
(\partial_{a_1},\partial_{a_3})^{n_{1,3}'-1}(\partial_{a_2},\partial_{a_3})^{n_{2,3}'} \nonumber\\(\partial_{a_1},p_2)^{s_1-m_1'}(\partial_{a_2},p_3)^{s_2-m_2'}(\partial_{a_3},p_1)^{s_3-m_3'+1}
\label{5.46}
\end{eqnarray}
Comparison with (5.31), (5.34) yields
\begin{eqnarray}
n_{1,2} = n_{1,2}',\, n_{1,3} = n_{1,3}'-1,\, n_{2,3} = n_{2,3}'; \qquad\qquad \nonumber\\
m_1 = m_1'-1,\, m_2 = m_2' ,\, m_3 = m_3'-1;\qquad\qquad \nonumber\\
s_1-m_1-1 = s_1-m_1',\, s_2-m_2 = s_2-m_2',\, s_3-m_3 = s_3-m_3' +1
\label{5.47}
\end{eqnarray}
The term (II) is
\begin{eqnarray}
(II):\qquad\qquad -n_{1,2}' (\partial_{a_1},\partial_{a_2})^{n_{1,2}'-1}(\partial_{a_1},\partial_{a_3})^{n_{1,3}'}(\partial_{a_2},\partial_{a_3})^{n_{2,3}'}
\nonumber\\
(\partial_{a_1},p_2)^{s_1-m_1'} (\partial_{a_2},p_3)^{s_2-m_2'+1}(\partial_{a_3},p_1)^{s_3-m_3'}
\label{5.48}
\end{eqnarray}
Comparison with (5.31), (5.34) yields now
\begin{eqnarray}
n_{1,2} = n_{1,2}'-1,\,n_{1,3} = n_{1,3}',\, n_{2,3} = n_{2,3}'; \qquad\quad \nonumber\\
m_1 = m_1'-1,\, m_2 = m_2' -1,\, m_3 = m_3' \qquad\qquad \nonumber\\
s_1-m_1 -1 = s_1-m_1',\, s_2-m_2 = s_2-m_2'+1,\, s_3-m_3 = s_3 - m_3'
\label{5.49}
\end{eqnarray}
All these relations are in both cases consistent in the sense
\begin{equation}
\sum_{i,j} n_{i,j} = \sum_{i,j}n_{i,j}'-1,\, \sum_{i} m_{i} = \sum_{i} m_{i}' -2
\label{5.50}
\end{equation}

The replacements
\begin{equation}
\{n'_{i,j}\}   \rightarrow \{n_{i,j}\}, \quad \{m_{i}'\} \rightarrow  \{m_{i}\}
\label{5.51}
\end{equation}
in the first line of (5.22) yield
\begin{eqnarray}
{\frac{M_{0}}{2}+1 \choose \{n_{i,j'}\}} \rightarrow (s_3+1)\, C^{s_1,s_2,s_3}_{n_1,n_2,n_3}\qquad\qquad \nonumber\\
{M_{0} +2 \choose \{m_{i}'\}}^{-1}\prod_{j=1}^{3} {s_{j}\choose m_{j}'} = [(2s_3+2)!]^{-1}\prod_{j=1}^{3}\frac{s_{j}!}{(s_{j}-m_{j}')!}
\label{5.52}
\end{eqnarray}
and for the whole first line in (5.26)
\begin{equation}
-\frac{1}{2} C^{s_1,s_2,s_3}_{n_1,n_2,n_3} [(2s_3)!(\Delta_{0}+2)!]^{-1}\prod_{i=1}^{3} \frac{s_{i}!}{(s_{i}-m_{i}')!}
\label{5.53}
\end{equation}
From (5.27) we find that the summations lead to the factors
\begin{equation}
(s_2-m_2'+1)(s_3-m_3'+1)
\label{5.54}
\end{equation}
which for the case I equals
\begin{equation}
(I): \qquad =(s_2-m_2+1)(s_3-m_3)
\label{5.55}
\end{equation}
and for the case II is
\begin{equation}
(II): \qquad =(s_2-m_2)(s_3-m_3+1)
\label{5.56}
\end{equation}
These have to be subtracted from each other
\begin{equation}
(I) - (II): \qquad = -(s_2-m_2) + (s_3-m_3)
\label{5.57}
\end{equation}
After multiplication with $1/2$ we subtract this quantity from the last line in (5.38) and obtain
\begin{equation}
\frac{1}{2}(s_1-m_1)
\label{5.58}
\end{equation}
thus getting the expected result $\frac{1}{2}n_1 =\frac{1}{2}(s_1-m_1)$ in (5.31).

\section{The quartic Noetherian}
\setcounter{equation}{0}
 In the case of $N$ = 4 there are three inequivalent possibilities to contract the currents, which lead each to gauge invariant expressions. Only their sum with fixed special values for the three coupling constants exhibits the desired Bose symmetry. We denote the three different contraction schemes
\begin{eqnarray}
...s_1 \leftrightarrow s_2 \leftrightarrow s_3 \leftrightarrow s_4 \leftrightarrow s_1... \nonumber\\
...s_1 \leftrightarrow s_3 \leftrightarrow s_2 \leftrightarrow s_4 \leftrightarrow s_1... \nonumber\\
...s_1 \leftrightarrow s_3 \leftrightarrow s_4 \leftrightarrow s_2 \leftrightarrow s_1...
\label{6.1}
\end{eqnarray}
by $s$-,$t$-, and $u$-channel, respectively. We study only the $s$-channel expression, the other ones are obtained by corresponding permutations. For the leading terms we neglect once again the divergence terms
\begin{equation}
(a_{i}, p_{i}) = 0
\label{6.2}
\end{equation}

\subsection{The case $n=0$}
In this case $n=0$ we extract the factors
\begin{equation}
{\frac{M}{2}\choose \{n_{i,j}\}}{M \choose \{m_{k}\}}^{-1}\prod_{\ell=1}^{4} {s_{\ell} \choose m_{\ell}}
\label{6.3}
\end{equation}
so that from (3.14) remain as relevant factors
\begin{eqnarray}
\sum_{partitions \,\{\pi_{\ell,\alpha} \}} \frac{(\prod_{\alpha=1}^{3} \zeta_{\alpha}!)}{(\sum_{\alpha =1}^{3}\zeta_{\alpha} +3)!}
\prod_{\ell=1}^{4}(-1)^{\eta_{\ell}}{s_{\ell}-m_{\ell} \choose k_{\ell},\{\pi_{\ell,\alpha}\}}\nonumber\\
(a_{\ell},w_{\ell-1})^{k_{\ell}}\prod_{\alpha = 1}^{3} (a_{\ell},w_{\alpha})^{\pi_{\ell,\alpha}}\qquad\qquad
\label{6.4}
\end{eqnarray}
By (6.2) this implies the following restrictions on the parameters $k_{\ell},\pi_{\ell,\alpha}$
\begin{equation}
k_1 = k_4 = 0,\qquad \pi_{1,1} = \pi_{4,3} = 0
\label{6.5}
\end{equation}
and the Fourier transformed differential operator part of (3.14) is
\begin{eqnarray}
\prod_{i,\alpha}(a_{i},w_{\alpha})^{\rho_{i,\alpha}} = \qquad\qquad\qquad \nonumber\\
(a_1,w_2)^{\rho_{1,2}}(a_1,w_3)^{\rho_{1,3}}(a_2,w_1)^{\rho_{2,1}}(a_2,w_3)^{\rho_{2,3}}\nonumber\\
(a_3,w_1)^{\rho_{3,1}}(a_3,w_2)^{\rho_{3,2}}(a_4,w_1)^{\rho_{4,1}}(a_4,w_2)^{\rho_{4,2}}
\label{6.6}
\end{eqnarray}
where
\begin{equation}
\rho_{2,1} = \pi_{2,1}+ \pi_{2,2} +k_2, \qquad \rho_{3,2} = \pi_{3,2}+\pi_{3,3} +k_3
\label{6.7}
\end{equation}
whereas for all other cases
\begin{equation}
\rho_{i,j} = \pi_{i,j}
\label{6.8}
\end{equation}
Moreover we note that
\begin{equation}
\sum_{i,j} \rho_{i,j} = S - M = \Delta, \, \sum_{i,j} \pi_{i,j} = S-M-K,\, K = k_2+k_3
\label{6.9}
\end{equation}
Now we consider the $\rho_{i,j}$ as fixed, leaving as variable parameters, say
\begin{equation}
k_2,\quad k_3, \quad \pi_{2,2},\quad \pi_{3,3}
\label{6.10}
\end{equation}
We have to sum over these.

Next we express the quantities $\zeta _{\alpha}$ by these now fixed variables $\rho_{i,j}$, respectively the variable parameters
(6.10)
\begin{eqnarray}
\zeta_1 = \rho_{2,1}+\rho_{3,1} + \rho_{4,1} -\pi_{2,2} -k_2= \textbf{R}_1-\pi_{2,2} -k_2 \nonumber\\
\zeta_2 = \rho_{1,2} +\rho_{3,2} +\rho_{4,2} +\pi_{2,2} -\pi_{3,3}-k_3 = \textbf{R}_2 +\pi_{2,2}-\pi_{3,3} -k_3\nonumber\\
\zeta_3 = \rho_{1,3} +\rho_{2,3} +\pi_{3,3} = \textbf{R}_3+\pi_{3,3}
\label{6.11}
\end{eqnarray}
We use the shorthands
\begin{equation}
 s_{i}-m_{i} = \nu_{i}
\label{6.12}
\end{equation}
which allows us to express
\begin{equation}
\sum_{\alpha} \zeta_{\alpha}   = \Delta -k_2 -k_3
\label{6.13}
\end{equation}
and to present (6.4) in the form
\begin{eqnarray}
\frac{s_1 !s_2 !s_3 !s_4 !}{M!}\sum_{partitions \{n_{i,j}\}} \sum_{\{\rho_{i,j}\}} {\frac{M}{2}\choose \{n_{i,j}\}}\quad \{
\frac{1}{\prod_{i=1}^{4}\nu_{i}!} \qquad\qquad\qquad\qquad \nonumber\\  \sum_{k_2,k_3,\pi_{2,2},\pi_{3,3}}(-1)^{k_2+k_3} \frac{(\textbf{R}_1-\pi_{2,2} -k_2)!(\textbf{R}_2+\pi_{2,2}-\pi_{3,3}-k_3)!
(\textbf{R}_3 +\pi_{3,3})!}{(\Delta-k_2-k_3 +3)!}\times\ \nonumber\\
{\nu_1 \choose \rho_{1,2},\rho_{1,3}}{\nu_2 \choose k_2,\rho_{2,1}-\pi_{2,2}-k_2,\pi_{2,2}, \rho_{2,3}}{\nu_3 \choose  k_3,\rho_{3,1},\rho_{3,2}-\pi_{3,3}-k_3, \pi_{3,3}}\times  \nonumber\\ {\nu_4 \choose \rho_{4,1}, \rho_{4,2}}\} [\prod_{i<j}(a_{i},a_{j})^{n_{i,j}}]
[\prod_{\ell,\alpha}(a_{\ell},w_{\alpha})^{\rho_{\ell,\alpha}}] \qquad\qquad\qquad\qquad
\label{6.14}
\end{eqnarray}
We recognize that the two quadrinomial coefficients depend on the free parameters $\pi_{2,2}, \pi_{3,3}, k_2, k_3$.
We denote this function within the curly bracket of (6.14) as
\begin{equation}
C_{\rho_{1,2},\rho_{1,3},\rho_{2,1},\rho_{2,3},\rho_{3,1},\rho_{3,2},\rho_{4,1},\rho_{4,2}}^{\nu_1,\nu_2,\nu_3,\nu_4}
\label{6.15}
\end{equation}

The summations over $k_2,k_3, \pi_{2,2},\pi_{3,3}$ can be performed by reduction to Gaussian series $_2F_1(1)$ for the first three summations and the fourth to a finite generalized hypergeometric series $_3F_2(1)$ of $\rho_{3,2}+1$ terms.
The result is
\begin{eqnarray}
C^{\nu_1,\nu_2,\nu_3,\nu_4}_{\rho_{1,2},\rho_{1,3},\rho_{2,1},\rho_{2,3},\rho_{3,1},\rho_{3,2},\rho_{4,1},\rho_{4,2}}\qquad\qquad\qquad\nonumber\\
=\frac{(-1)^{\rho_{2,1}+\rho_{3,2}}}{\prod_{i}\nu_{i}!} {\nu_1\choose \rho_{1,2},\rho_{1,3}}{\nu_2 \choose \rho_{2,1},\rho_{2,3}}{\nu_3 \choose \rho_{3,1},
\rho_{3,2}}{\nu_4 \choose \rho_{4,1},\rho_{4,2}}\nonumber\\
\times\frac{(\textbf{R}_1-\rho_{2,1})!(\textbf{R}_2-\rho_{3,2})! \textbf{R}_3!}{(\Delta+3)!}(\Delta-\textbf{R}_2+3)_{\rho_{3,2}} (\Delta-\textbf{R}_1-\textbf{R}_2 +2)_{\rho_{2,1}} \nonumber\\
\times_3F_2(-\rho_{3,2}, \textbf{R}_3+1,\Delta - \textbf{R}_1-\textbf{R}_2 +\rho_{2,1} +2;\Delta -\textbf{R}_1-\textbf{R}_2+2,
\Delta-\textbf{R}_2+3; 1) \nonumber\\ \quad
\label{6.16}
\end{eqnarray}
Here we have not made use of the fact that $\textbf{R}_1+\textbf{R}_2+\textbf{R}_3 = \Delta$, but we have kept these four
parameters as independent. This turns out to be very practical when we treat the cases $n>0$.

\subsection{The cases $n \geq 1$}
In this case we introduce the factor $Z$ which in the notation using the momenta $\{w_{i}\}$ is
\begin{equation}
Z =  \sum_{i=1}^{3}\tau_{i+1}(1-\tau_{i+1})w_{i}^2 -2\sum_{i\leq j}\tau_{i+1}\tau_{j+1}(w_{i},w_{j})
\label{6.17}
\end{equation}
which can be expressed by the basis described in the Appendix
\begin{equation}
Z = \sum_{i=1}^{4} A_{i}(w_{i}-w_{i-1})^2 + \sum_{1\leq j<k\leq 3}B_{j,k}(w_{j},w_{k}),
\label{6.18}
\end{equation}
where the coefficients $A_{i}, B_{j,k}$ (8.1) are functions of $\tau_2, \tau_3, \tau_4$ specified in (8.5).

Any n'th power of $Z$ can be multinomially expanded
\begin{eqnarray}
Z^{n} = \sum_{n_1,n_2,\{\lambda_{i}\},\{\mu_{j,k}\}}{n\choose n_1,n_2}{n_1 \choose \lambda_1,\lambda_2,\lambda_3,\lambda_4}
{n_2 \choose \mu_{1,2},\mu_{1,3},\mu_{2,3}}\nonumber\\
((w_1)^2)^{\lambda_1}((w_2-w_1)^2)^{\lambda_2}((w_3-w_2)^2)^{\lambda_3}((w_3)^2)^{\lambda_4}\qquad \nonumber\\(w_1,w_2)^{\mu_{1,2}} (w_1,w_3)^{\mu_{1,3}}(w_2,w_3)^{\mu_{2,3}}
A_1^{\lambda_1}A_2^{\lambda_2}A_3^{\lambda_3}A_4^{\lambda_4}\quad B_{1,2}^{\mu_{1,2}+\mu_{2,3}}B_{1,3}^{\mu_{1,3}}
\label{6.19}
\end{eqnarray}
The product over the $A$ and $B$ coefficients has to be integrated with the measure
\begin{equation}
\int (\prod_{j=1}^{3}d\tau_{j+1}\theta(\tau_{j+1})) \prod_{\alpha=1}^{3}\tau_{\alpha+1}^{\zeta_{\alpha}}
\label{6.20}
\end{equation}
where in analogy with (6.3),(6.4) we have
\begin{eqnarray}
\zeta_{\alpha} = \sum_{i=1}^{4}\pi_{i,\alpha}, \quad \sum_{\alpha=1}^{3}\zeta_{\alpha} = \Delta_{n} -k_2-k_3, \nonumber\\
\Delta_{n} =  \Delta -2n \qquad\qquad
\label{6.21}
\end{eqnarray}
and
\begin{equation}
\sum_{i<j}n_{i,j}' = \frac{1}{2}M_{n} = \frac{1}{2}M +n, \quad \sum_{k} m_{k}' = M_{n} = M+2n
\label{6.22}
\end{equation}
The terms $n_1 \neq 0$ belong to the kernel, whereas those with $n_1 = 0$ contribute to the leading terms. We will continue to discuss only these.

In this case the number of derivatives $\Delta$ restricts $n$ to
\begin{equation}
0 \leq n \leq \frac{1}{2}\Delta
\label{6.23}
\end{equation}
In such case we must integrate
\begin{eqnarray}
\sum_{\mu_{i,j}} {n\choose\mu_{1,2},\mu_{1,3},\mu_{2,3}}(w_1,w_2)^{\mu_{1,2}}(w_1,w_3)^{\mu_{1,3}}(w_2,w_3)^{\mu_{2,3}}\,B_{1,2}^{\mu_{1,2}+\mu_{2,3}}B_{1,3}^{\mu_{1,3}}=\nonumber\\
\sum_{\mu_{1,3}}{n \choose \mu_{1,3}}(w_1,w_3)^{\mu_{1,3}} ((w_1,w_2)+(w_2,w_3))^{\mu_{1,2}+\mu_{2,3}}\,B_{1,2}^{\mu_{1,2}+\mu_{2,3}}B_{1,3}^{\mu_{1,3}}
\label{6.24}
\end{eqnarray}
In the sequel we denote
\begin{equation}
\mu_{1,2}+\mu_{2,3} = \mu_{1,2,3} = n - \mu_{1,3}
\label{6.25}
\end{equation}

The integration is over the measure (6.20) and after introduction of the new variable $\tau_1 = 1-\tau_2-\tau_3-\tau_4$ yields $(-2)^{\mu_{1,3}}$ times (see (3.5))
\begin{eqnarray}
\int d\tau_1 d\tau_2 d\tau_3 d\tau_4 \delta(1-\tau_1-\tau_2-\tau_3-\tau_4)\tau_1^{\mu_{1,2,3}}\tau_{2}^{\zeta_1+ \mu_{1,3}}
\tau_{3}^{\zeta_2+\mu_{1,2,3}}\tau_4^{\zeta_3+\mu_{1,3}} \nonumber\\ =
\frac{\mu_{1,2,3}!(\zeta_1+\mu_{1,3})! (\zeta_2+\mu_{1,2,3})!(\zeta_3+\mu_{1,3})!}{(\Delta -k_2-k_3+3)!} \qquad\qquad
\label{6.26}
\end{eqnarray}
where the denominator is as usual independent of $n$. Taking into account the factor $(-1)^{n}/n!$ from the residue of the gamma function (2.28) we obtain a finite sum
\begin{eqnarray}
(-1)^{n}\sum_{\mu_{1,3}= 0}^{n}  \frac{(\zeta_1+\mu_{1,3})! (\zeta_2+n-\mu_{1,3})!(\zeta_3+\mu_{1,3})!}{\mu_{1,3}!
(\Delta-k_2-k_3+3)!}\qquad\qquad\qquad\nonumber\\  (-2(w_1,w_3))^{\mu_{1,3}}[(w_1,w_2)+(w_2,w_3)]^{n -\mu_{1,3}}
\label{6.27}
\end{eqnarray}
of the type of a (cut off at $\mu_{1,3} = n$) Gaussian hypergeometric series.
From now on we simplify the notations a little bit by defining
\begin{equation}
\mu_{1,3} = \mu, \quad\mu_{1,2,3} = n-\mu
\label{6.28}
\end{equation}

For arbitrary $n$ and using as before $M_{n} = M +2n$ and corresponding labels $\{n_{i,j}', m_{k}'\}$ we obtain as starting point instead of (3.14)
\begin{eqnarray}
\Gamma(\frac{M_{n}}{2})2^{\frac{M_{n}}{2}}\sum_{partitions\,\{n_{i,j}'\}}{\frac{M_{n}}{2} \choose \{n_{i,j}'\}}{M_{n} \choose \{m_{k}'\}}^{-1}(\prod_{i<j}(a_{i},a_{j})^{n_{i,j}'})\qquad\qquad\qquad \nonumber\\
\sum_{partitions \,\{\pi_{\ell,\alpha}\}, k_{\ell}}\frac{(\zeta_1'+\mu)!(\zeta_2'+n-\mu)!(\zeta_3'+\mu)!}{\mu! (\Delta-k_2-k_3+3)!}\qquad\qquad\qquad\qquad
\nonumber\\(1/2(a_{\ell}, w_{\ell}+w_{\ell-1}))^{k_{\ell}}(\prod_{\ell,\alpha}(a_{\ell},w_{\alpha})^{\pi_{\ell,\alpha}}) [-2(w_1,w_3)]^{\mu}[(w_1,w_2)+(w_2,w_3)]^{n-\mu}\qquad
\label{6.29}
\end{eqnarray}
Here the parameters $k_2,k_3,\pi_{2,2},\pi_{3,3}$ remain free and must be summed over.

Now we proceed exactly as before and introduce a modified coefficient function $C$ depending on $n, \mu, \nu_{i}'=s_{i}-m_{i}'$
\begin{eqnarray}
C_{\rho_{1,2},\rho_{1,3},\rho_{2,1},\rho_{2,3},\rho_{3,1},\rho_{3,2},\rho_{4,1},\rho_{4,2}}^{\nu_1',\nu_2',\nu_3',\nu_4'}
[n,\mu]  =
\frac{1}{\prod_{i=1}^{4} \nu_{i}'}\sum_{k_2,k_3,\pi_{2,2},\pi_{3,3}} (-1)^{k_2+k_3}\qquad\qquad\nonumber\\ \frac{(\textbf{R}_1+\mu-\pi_{2,2}-k_2)!(\textbf{R}_2+n -\mu+\pi_{2,2}-\pi_{3,3}-k_3)!(\textbf{R}_3 +\mu+\pi_{3,3})!}{\mu!(\Delta-k_2-k_3+3)}\nonumber\\ \times
{\nu_1' \choose \rho_{1,2},\rho_{1,3}}{\nu_2' \choose k_2,\rho_{2,1}-\pi_{2,2}-k_2,\pi_{2,2},\rho_{2,3}}{\nu_3' \choose
k_3,\rho_{3,1},\rho_{3,2}-\pi_{3,3}-k_3,\pi_{3,3}}\qquad\nonumber\\\times{\nu_4' \choose \rho_{4,1},\rho_{4,2}}\qquad\qquad\qquad
\qquad\qquad
\label{6.30}
\end{eqnarray}
It is obvious that for general $n,\mu$ we have only to replace in (6.14) and (6.16)
\begin{equation}
\textbf{R}_1 \Rightarrow \textbf{R}_1+\mu,\, \textbf{R}_2 \Rightarrow \textbf{R}_2 + n-\mu, \,\textbf{R}_3 \Rightarrow \textbf{R}_3 + \mu
\label{6.31}
\end{equation}
whereas we keep $\Delta$ fixed, what is allowed if these four parameters are considered as independent. So instead of (6.16) we get now
\begin{eqnarray}
C_{\rho_{1,2},\rho_{1,3},\rho_{2,1},\rho_{2,3},\rho_{3,1},\rho_{3,2}\rho_{4,1},\rho_{4,2}}^{\nu_1',\nu_2',\nu_3',\nu_4'}
[n,\mu] = \qquad\qquad\qquad\qquad\nonumber\\
\frac{(-1)^{\rho_{2,1}+\rho_{3,2}}}{\prod_{i}\nu_{i}'!}{\nu_1'\choose \rho_{1,2},\rho_{1,3}}{\nu_2' \choose \rho_{2,1},\rho_{2,3}}{\nu_3' \choose \rho_{3,1},\rho_{3,2}}{\nu_4' \choose \rho_{4,1},\rho_{4,2}}\times\qquad\qquad \nonumber\\
\frac{(\textbf{R}_1 +\mu-\rho_{2,1})!(\textbf{R}_2 +n-\mu -\rho_{3,2})!(\textbf{R}_3+\mu)!}{\mu!(\Delta+3)!}\times \nonumber\\(\Delta-\textbf{R}_2-n+\mu+3)_{\rho_{3,2}}(\Delta -\textbf{R}_1-\textbf{R}_2 -n
+2)_{\rho_{2,1}} \times \nonumber\\
_3F_2(-\rho_{3,2},\textbf{R}_3 + \mu +1, \Delta -\textbf{R}_1 -\textbf{R}_2 -n +\rho_{2,1} +2;\nonumber\\ \Delta-\textbf{R}_1 -\textbf{R}_2 -n+2, \Delta -\textbf{R}_2-n+\mu +3;1)
\label{6.32}
\end{eqnarray}
where $0\leq \mu \leq n\leq \frac{\Delta}{2}$. Thus for $n=0$ (6.32) reduces to (6.16).

Finally we give the result for all leading terms of the Lagrangian where the $m_{k}'$ are determined by the $n_{i,j}'$
through (2.44)
\begin{eqnarray}
\sum_{n=0}^{\frac{\Delta}{2}}(-1)^{n}\Gamma(\frac{M_{n}}{2})2^{\frac{M_{n}}{2}}\sum_{partitions\,\{n_{i,j}'\}}
{\frac{M_{n}}{2} \choose \{n_{i,j}'\}}{M_{n} \choose \{m_{k}'\}}^{-1}\sum_{\mu=0}^{n}\sum_{\{\rho_{i,j}\}}\nonumber\\
C_{\rho_{1,2},\rho_{1,3},\rho_{2,1},\rho_{2,3},\rho_{3,1},\rho_{3,2},\rho_{4,1},\rho_{4,2}}^{s_1-m_1',s_2-m_2',s_3-m_3',
s_4-m_4'}[n,\mu] \prod_{\ell,\alpha}(a_{\ell},w_{\alpha})^{\rho_{\ell,\alpha}}\prod_{i<j}(a_{i},a_{j})^{n_{i,j}'}\nonumber\\
\qquad[-2(w_1,w_3)]^{\mu}(w_1+w_3,w_2)^{n-\mu}\qquad\qquad\qquad
\label{6.33}
\end{eqnarray}

\section{Conclusion}
For the $N=2$ Noetherian we have obtained an expression which agreed with Fronsdal's free Lagrangian up to the trace terms,
but the latter ones where wrong. The $B$-parameter had to be chosen zero in order not to produce $D$-dependent coefficients
of the trace terms. Since the $N=2$ Noetherian is conformally invariant, the free Lagrangian cannot be.

For the case $N=3$ we reproduced the leading terms and some of the divergence terms (deDonder terms for vanishing trace field) of the known cubic interaction Lagrangian \cite{MMR2, MMR3, MMR4}. We know from \cite{MMR3} that these terms are sufficient to reproduce the whole interaction Lagrangian uniquely ("Proposition 1"), so that our present result agrees completely with our former result for the cubic interaction, and this interaction is now known to be conformally invariant.

For the case $N=4$ three versions (s-, t-, and u-channel) of the Noether potential exist. Bose symmetry requires a symmetric sum of them to be the complete expression. We derived the leading terms of the s-channel version. It would be interesting whether these incomplete results can be completed by a version of "Proposition 1", perhaps provable from conformal invariance.

The algorithm presented here is also applicable to analogous problems on spaces with a conformal structure such as AdS spaces.

Acknowledgement: The author thanks Ruben Manvelyan and Karapet Mkrtchyan for clarifying discussions and helpful advices, and the organizers of the Third Workshop on String Theory, July 2011, at the Witwatersrand University in Johannesburg, where a preliminary version of this work was presented.

\section{Appendix: A basis of bilinear forms for the momenta $w_{i}$}
\setcounter{equation}{0}
In our analysis the Laplacians
\begin{equation}
(w_{i}-w_{i-1})^{2},\quad i\in \{1,2,...N\}
\label{A1}
\end{equation}
play a special role for the selection of the kernel which vanishes on shell. Therefore we make the following ansatz
for an arbitrary bilinear form
\begin{equation}
\sum_{i=1}^{N} A_{i} (w_{i}-w_{i-1})^{2} + \sum_{1\leq i < j \leq N-1} B_{i,j} (w_{i},w_{j})
\label{A2}
\end{equation}
The dimension of this space of bilinear forms, namely $N+{N-1 \choose 2}$ is, however, one too big, since by momentum conservation we may set $A_{N}=0$. On the other hand we need all Laplacians. Therefore we try to imply a constraint on the coefficients
$B_{i,j}$ which is noncontradictory, e.g. the symmetric constraint
\begin{equation}
B_{1,2} = B_{N-2,N-1}
\label{A3}
\end{equation}
Applying this ansatz to $Z$ (6.17), we obtain the equations
\begin{eqnarray}
A_1+A_2 = \tau_2(1-\tau_2) \nonumber\\
A_2+A_3 = \tau_3(1-\tau_3) \nonumber\\
A_3+A_4 = \tau_4(1-\tau_4) \nonumber\\
B_{1,3} = -2\tau_2\tau_4 \nonumber\\
B_{1,2}-2A_2 = -2 \tau_2\tau_3 \nonumber\\
B_{2,3}-2A_3 = -2 \tau_3\tau_4
\label{A4}
\end{eqnarray}
This leads to the following solutions
\begin{eqnarray}
A_1= \tau_2(1-\tau_2) -\frac{1}{2}\tau_3 (1+\tau_2-\tau_3-\tau_4) \nonumber\\
A_2= \frac{1}{2}\tau_3 (1+\tau_2-\tau_3-\tau_4) \qquad\nonumber\\
A_3 =\frac{1}{2}\tau_3 (1-\tau_2-\tau_3+\tau_4) \qquad\nonumber\\
A_4 =\tau_4(1-\tau_4) -\frac{1}{2}\tau_3(1-\tau_2-\tau_3+\tau_4)\nonumber\\
B_{1,2} = B_{2,3} = \tau_3(1-\tau_2-\tau_3-\tau_4)\nonumber\\
B_{1,3} = -2\tau_2\tau_4 \qquad\qquad
\label{A5}
\end{eqnarray}

\end{document}